\documentclass[prd,superscriptaddress,nofootinbib,twocolumn,twoside,preprintnumbers]{revtex4-2}

\usepackage[utf8]{inputenc}
\usepackage{setspace}
\usepackage[colorlinks]{hyperref}
\hypersetup{
    colorlinks = true,
    citecolor  = red,
	linkcolor  = blue
}

\usepackage{bbm}
\usepackage{amsfonts}
\usepackage{mathrsfs}

\usepackage{bbold}
\usepackage{amsmath,amssymb}
\usepackage[english]{babel}
\usepackage{siunitx}
\usepackage{slashed}

\usepackage{epsfig}
\usepackage{graphicx}               
\usepackage{url}
\usepackage{float}
\usepackage{soul}
\usepackage{pstricks}
\usepackage{color}
\usepackage{multirow}

\usepackage{threeparttable}
\usepackage[version=4]{mhchem}

\usepackage{colortbl}
\usepackage{booktabs}
\usepackage{fontawesome}

\newcommand{\MM}{\mathcal{M}}

\def\ve{v_\text{e}}
\def\vve{\vect{v}_\text{e}}
\def\vveh{\vect{\hat v}_\text{e}}
\def\vesc{v_\text{esc}}
\def\vect#1{\boldsymbol{#1}}

\def\ie{{\it i.e.}}
\def\eg{{\it e.g.}}

\def\codelink{https://github.com/tanner-trickle/dm-phonon-scatter}
\def\demolink{https://demo-phonon-web-app.herokuapp.com}

\begin{document}

\title{Directional Detectability of Dark Matter With Single Phonon Excitations:\\ Target Comparison}

\author{Ahmet Coskuner}
\affiliation{Department of Physics, University of California, Berkeley, CA 94720, USA}
\affiliation{Theoretical Physics Group, Lawrence Berkeley National Laboratory, Berkeley, CA 94720, USA}

\author{Tanner Trickle}
\author{Zhengkang Zhang}
\author{Kathryn M.~Zurek}
\affiliation{Walter Burke Institute for Theoretical Physics, California Institute of Technology, Pasadena, CA 91125, USA}

\preprint{CALT-TH-2021-008}

\begin{abstract}
    Single phonon excitations are sensitive probes of light dark matter in the keV-GeV mass window. For anisotropic target materials, the signal depends on the direction of the incoming dark matter wind and exhibits a daily modulation. We discuss in detail the various sources of anisotropy, and carry out a comparative study of 26 crystal targets, focused on sub-MeV dark matter benchmarks. We compute the modulation reach for the most promising targets, corresponding to the cross section where the daily modulation can be observed for a given exposure, which allows us to combine the strength of DM-phonon couplings and the amplitude of daily modulation.  We highlight Al$_2$O$_3$ (sapphire), CaWO$_4$ and h-BN (hexagonal boron nitride) as the best polar materials for recovering a daily modulation signal, which feature $\mathcal{O}(1$\,--\,$100)$\% variations of detection rates throughout the day, depending on the dark matter mass and interaction.  The directional nature of single phonon excitations offers a useful handle to mitigate backgrounds, which is crucial for fully realizing the discovery potential of near future experiments.
\end{abstract}

\maketitle

\section{Introduction}
\label{sec:intro}

If the cold dark matter (DM) in the universe consists of new particles, they must interact very weakly with the Standard Model. Directly detecting these feeble interactions in a laboratory requires extraordinarily sensitive devices. Traditional direct detection experiments (\eg\ ANAIS~\cite{Amare:2019ncj}, CRESST~\cite{Probst:2002qb, Angloher:2018fcs, Kluck:2020bdm}, DAMA/LIBRA~\cite{Bernabei:2020mon}, DAMIC~\cite{Aguilar-Arevalo:2020oii}, DarkSide-50~\cite{Agnes:2018ves}, DM-Ice~\cite{Jo:2016qql}, KIMS~\cite{Kim:2015prm}, LUX~\cite{Akerib:2019zrt,Akerib:2020gzy}, SABRE~\cite{Antonello:2018fvx}, SuperCDMS~\cite{Agnese:2018gze,Alkhatib:2020slm}, and Xenon1T~\cite{Aprile:2019xxb}), based on nuclear recoil, are gradually improving their sensitivity and closing the open parameter space before reaching the irreducible solar and atmospheric neutrino background. However, these experiments are fundamentally limited in the DM mass, $m_\chi$, they can probe. When the DM scatters off a nucleus at rest, the energy deposited, $\omega$, is limited by $\omega \lesssim m_\chi^2 v^2/m_N$, with $v\sim 10^{-3}$, and vanishes quickly as the DM mass decreases below the nucleus mass $m_N$.

This limitation in DM mass is typically not problematic in the search for the prototypical weakly interacting massive particle (WIMP) which produces the DM abundance through freeze-out, as $m_\chi \lesssim$ GeV would both be overabundant and be in tension with indirect detection bounds on DM annihilation rates. However, many other theoretically motivated explanations of the origin of DM such as freeze-in~\cite{Hall:2009bx, Bernal:2017kxu}, hidden sector DM \cite{Strassler:2006im,ArkaniHamed:2008qp, Cheung:2009qd, Morrissey:2009ur}, asymmetric DM~\cite{Kaplan:2009ag, Cohen:2010kn, Petraki:2013wwa, Zurek:2013wia}, and strong self interactions \cite{Hochberg:2014dra, Hochberg:2014kqa}, allow for DM ligher than a GeV and therefore should be searched for by means other than nuclear recoil.

In the pursuit of sub-GeV DM, several new experimental concepts have been proposed. These include electron excitations in a variety of target systems~\cite{Essig:2011nj,Graham:2012su,Lee:2015qva,Essig:2015cda,Derenzo:2016fse, Hochberg:2016ntt, Essig:2017kqs,Kurinsky:2019pgb,Blanco:2019lrf,Catena:2019gfa,Settimo:2020cbq,Barak:2020fql,Amaral:2020ryn} for DM with mass above an MeV, while single (primary) phonon excitations~\cite{Schutz:2016tid, Knapen:2016cue, Acanfora:2019con, Caputo:2019cyg, Baym:2020uos, Caputo:2020sys, Knapen:2017ekk, Griffin:2018bjn, Trickle:2019nya, Griffin:2019mvc,Campbell-Deem:2019hdx,Griffin:2020lgd, Trickle:2020oki,Kahn:2020fef,Knapen:2020aky}, with energies up to $\mathcal{O}(100)$ meV, have been shown to be especially sensitive to a wide range of DM models with masses down to a keV. This coupled with the fact that detector energy thresholds are approaching the $\mathcal{O}(100)$ meV range~\cite{Pyle:2015pya, Maris:2017xvi, Rothe:2018bnc, Colantoni:2020cet, Fink:2020noh} makes single phonon excitations an exciting avenue for DM direct detection. Phonons are quasiparticle vibration quanta which can exist in multiple states of matter, \eg\ as sound waves in liquids or superfluids and lattice vibrations in crystalline solids. Superfluid helium has been proposed~\cite{Schutz:2016tid} and studied as a light DM detector \cite{Knapen:2016cue,Acanfora:2019con, Caputo:2019cyg, Baym:2020uos, Caputo:2020sys} and an experiment is currently in the R\&D phase~\cite{tesseract}. Crystal targets have also been proposed \cite{Knapen:2017ekk} and studied extensively. Initial studies focused on \ce{GaAs} and \ce{Al2O3} (sapphire) targets \cite{Griffin:2018bjn}, and more recently this analysis has been extended to account for more general DM interactions \cite{Trickle:2019nya, Trickle:2020oki} and applied to a broader set of target materials \cite{Griffin:2019mvc}. Other targets have also been proposed individually, \eg\ \ce{SiC} \cite{Griffin:2020lgd}, and there has been work on understanding the signal from multi-phonon excitations \cite{Campbell-Deem:2019hdx,Kahn:2020fef,Knapen:2020aky}. Similar to superfluid helium, a DM detector using a crystal target with single phonon readout is also in the R\&D phase of development \cite{tesseract}. 

Most of the previous work has focused on calculating the theoretically predicted DM-phonon interaction strength. Equally important is to minimize the experimental background~\cite{Du:2020ldo}. This becomes easier when the DM scattering signal has unique properties which can distinguish it from backgrounds. For example, in experiments sensitive to nuclear recoil or electron excitations, the rate modulates annually due to the change in the DM velocity distribution in the Earth frame, as the Earth orbits around the Sun~\cite{Drukier:1986tm,Essig:2015cda,Lee:2015qva}.

In an experiment based on primary phonon readout, the DM scattering rate can have a larger and more unique signature: \textit{daily} modulation. As the Earth rotates about its own axis, the orientation of the detector relative to the DM wind changes. In a nuclear recoil experiment this does not have an effect since the interaction matrix element is independent of the direction of the DM velocity relative to the detector orientation --- an (unpolarized) nucleus is isotropic in its response. However, crystal targets can be highly anisotropic, which means that the amplitude of the response depends not only on the magnitude of the momentum transfer but also on its direction.  This can lead to a significant daily modulation. Moreover, since the modulation pattern depends on the crystal orientation, running an experiment with multiple detectors simultaneously with different orientations can further enhance the signal-to-noise ratio. This effect was studied for sapphire in Ref.~\cite{Griffin:2018bjn} (see also Refs.~\cite{Hochberg:2016ntt,Coskuner:2019odd,Geilhufe:2019ndy,Trickle:2019nya} for discussions of daily modulation in electron excitations). In this work, we expand the understanding of the daily modulation effect in single phonon excitations to a broader range of materials, including those targeted in Ref.~\cite{Griffin:2019mvc}. 

In particular, we highlight the following targets in the main text: \ce{Al2O3} (sapphire) and \ce{CaWO4}, which were already utilized for DM detection and have a significant daily modulation; \ce{SiO2}, which was shown to have a strong reach to several benchmark models; \ce{SiC}, which was proposed in Ref.~\cite{Griffin:2020lgd} (for which we choose the commercially available 4H polytype); and h-BN (hexagonal boron nitride), which is a highly anisotropic material. Among them, \ce{Al2O3} and \ce{CaWO4} have the best prospects overall in terms of daily modulation reach and experimental feasibility. 

Of the additional materials considered in Ref.~\cite{Griffin:2019mvc}, results for those with daily modulation larger than 1\% (for at least some DM masses where the material has substantial reach) are presented in an appendix.  We make available our code for the phonon rate calculation~\cite{dm-phonon-scatter}, and also publish an interactive webpage~\cite{demo} where results for all the materials presented in Ref.~\cite{Griffin:2019mvc} and in this paper, including reach curves, differential rates and daily modulation patterns, can be generated from our calculations. Twenty-six materials are initially included on the interactive webpage~\cite{demo}: \ce{Al2O3}, \ce{AlN}, h-BN, \ce{CaF2}, \ce{CaWO4}, \ce{CsI}, C (diamond), \ce{GaAs}, Ge, \ce{GaN}, \ce{GaSb}, \ce{InSb}, \ce{LiF}, \ce{MgF2}, \ce{MgO}, \ce{NaCl}, \ce{NaF}, \ce{NaI}, \ce{PbS}, \ce{PbSe}, \ce{PbTe}, Si, 4H-SiC, \ce{SiO2}, \ce{ZnO}, and \ce{ZnS}. 
This diverse set of materials (with some currently in use in nuclear recoil experiments, some proposed for light dark matter detection, and some others being promising polar crystals from theoretical considerations) aims to explore a wide range of possibilities with the hope of identifying broad theoretical features that could be implemented in a more practical experimental setup. Materials shown in-text (\ce{Al2O3}, \ce{SiO2}, \ce{SiC}, \ce{CaWO4} and h-BN) are the ones with the highest daily modulation in this list.

\section{Directional Detection With \\Single Phonon Excitations}
\label{sec:paths}

\subsection{Excitation Rate}
\label{subsec:ex_rate}

We begin by summarizing the formulae for single phonon excitation rates; see Refs.~\cite{Knapen:2017ekk,Griffin:2018bjn,Trickle:2019nya} for more details. For the scattering of a DM particle $\chi$ with mass $m_\chi$ and general spin-independent interactions, the rate per unit target mass takes the form
\begin{eqnarray}
    R(t) = \frac{1}{\rho_T} \frac{\rho_\chi}{m_\chi} &&\frac{\pi\overline\sigma_{\psi}}{\mu_{\chi\psi}^2} \int d^3 v \, f_\chi(\vect{v},t) \nonumber\\
&& \times \int\frac{d^3q}{(2\pi)^3} \,{\cal F}_\text{med}^2(q)\,S\bigl(\vect{q},\omega_{\vect{q}})  \, ,
    \label{eq:rate}
\end{eqnarray}
where $\vect{v}$ is the incoming DM's velocity, $\vect{q}$ is the momentum transferred to the target, $\rho_T$ is the target's mass density, and $\rho_\chi=0.4\,$GeV/cm$^3$ is the local DM density. $\overline\sigma_{\psi}$, with $\psi=n$ or $e$ (neutron or electron), is a reference cross section defined as
\begin{equation}
\overline\sigma_\psi \equiv \frac{\mu^2_{\chi \psi}}{\pi} \overline{|\MM_{\chi\psi}(q=q_0)|^2}\,,
\end{equation}
where $\mu_{\chi\psi}$ is the reduced mass, $\MM_{\chi\psi}$ is the vacuum matrix element for $\chi\psi\to\chi\psi$ scattering, and $q_0$ is a reference momentum transfer. We present the reach in terms of $\overline\sigma_\psi$, with $q_0=m_\chi v_0$ (where $v_0=230\,$km/s, the dispersion of DM's velocity distribution) for $\psi=n$ and $q_0=\alpha m_e$ for $\psi=e$. $f_\chi(\vect{v}, t)$ is the DM's velocity distribution in the lab frame, taken to be a truncated Maxwell-Boltzmann distribution, boosted by the time-dependent Earth velocity $\vve(t)$, as will be discussed in more detail in the next subsection. ${\cal F}_\text{med}(q)$ is the mediator form factor, which captures the $q$ dependence of the mediator propagator:
\begin{equation}
{\cal F}_\text{med}(q) =
\begin{cases}
1 & \text{(heavy mediator)} \,,\\
(q_0/q)^2 & \text{(light mediator)} \, .
\end{cases}
\label{eq:Fmed}
\end{equation}
Finally, $S(\vect{q},\omega)$ is the dynamic structure factor that encodes target response to DM scattering with momentum transfer $\vect{q}$ and energy transfer $\omega$, constrained by energy-momentum conservation to be
\begin{equation}
\omega_{\vect{q}} = \vect{q}\cdot\vect{v} -\frac{q^2}{2m_\chi}\,.
\end{equation}
Generally, one sums over a set of final states $f$ with energies $\omega_f$, and $S(\vect{q},\omega)$ takes the form
\begin{equation}
S(\vect{q},\omega) = \sum_f 2\pi\, \delta(\omega-\omega_f) \,S'_f(\vect{q}) \,.
\end{equation}
For single phonon excitations, we assume the target system is initially prepared in the ground state at zero temperature with no phonons, and sum over single phonon states labeled by branch $\nu$ and momentum $\vect{k}$ inside the first Brillouin zone (1BZ). Lattice momentum conservation dictates that $\vect{q} = \vect{k}+\vect{G}$, with $\vect{G}$ a reciprocal lattice vector. To find $\vect{k}$ and $\vect{G}$ from a given $\vect{q}$, we first find the reduced coordinates $(q_1, q_2, q_3)$ (\ie\ $\vect{q}=\sum_{i=1}^3 q_i \vect{b}_i$ with $\vect{b}_i$ the basis vectors of the reciprocal lattice), and then find the nearest point $(G_1, G_2, G_3)$ with $G_i\in\mathbb{Z}$. In this way, any $\vect{q}$ outside of the 1BZ is mapped to a $\vect{k}$ inside the 1BZ and a $\vect{G}$ vector. The sum over final states therefore only runs over the phonon branches, indexed by $\nu$,
\begin{equation}
S(\vect{q},\omega) = \sum_\nu 2\pi \delta(\omega-\omega_{\nu,\vect{k}}) \,S'_\nu(\vect{q}) \,.
\label{eq:S_sum_nu}
\end{equation}
As was shown in Refs.~\cite{Knapen:2017ekk,Griffin:2018bjn,Trickle:2019nya}, $S'_\nu$ can be written in terms of the phonon energies $\omega_{\nu\vect{k}}$, eigenvectors $\vect{\epsilon}_{\nu\vect{k}j}$ and an effective DM-ion couplings $\vect{Y}_j$ (with $j$ labeling the ions in the primitive cell):
\begin{equation}
S'_\nu(\vect{q}) = \frac{1}{2\Omega\,\omega_{\nu,\vect{k}}}\, \biggl|\sum_{j} \frac{e^{-W_j(\vect{q})}}{\sqrt{m_j}}\,e^{i\vect{G}\cdot \vect{x}_j^0} \bigl(\vect{Y}_j\cdot\vect{\epsilon}_{\nu,\vect{k},j}^*\bigr) \biggr|^2,
\label{eq:structure_phonon}
\end{equation}
where $\Omega$ is the volume of the primitive cell, and $m_j, \vect{x}_j^0$, and $W_j (\vect{q}) \equiv \frac{\Omega}{4m_j}\sum_\nu \int\frac{d^3k}{(2\pi)^3} \frac{|\vect{q}\cdot\vect{\epsilon}_{\nu,\vect{k},j}|^2}{\omega_{\nu,\vect{k}}}$ are the masses, equilibrium positions, and Debye-Waller factors of the ions, respectively. We obtain the material-specific force constants in the quadratic crystal potential and the equilibrium positions from density functional theory (DFT) calculations~\cite{Griffin:2018bjn, Griffin:2019mvc,phonondb}, and use the open-source phonon eigensystem solver \texttt{phonopy}~\cite{phonopy} to derive the values of $\omega_{\nu,\vect{k}}$, $\vect{\epsilon}_{\nu,\vect{k},j}$ for each material.

The DM-ion coupling vectors $\vect{Y}_j$ are DM model dependent. In our target comparison study in Sec.~\ref{sec:target_comparison}, we will focus on two sets of benchmark models, with a light dark photon mediator and a heavy or light hadrophilic scalar mediator, respectively. These are the same models considered in Ref.~\cite{Griffin:2019mvc}, for which $\vect{Y}_j$ are given by
\begin{equation}
    \label{eq:Yj}
\vect{Y}_j = 
\begin{cases}
-  \dfrac{\vect{q}\cdot\vect{Z}_j^\star}{\vect{\hat q} \cdot \vect{\varepsilon}_\infty \cdot \vect{\hat q}} & \text{(dark photon med.)},\\[12pt]
\vect{q}\,A_j\, F_{N_j}(q) & \text{(hadrophilic scalar med.)}.
\end{cases}
\end{equation}
Here $\vect{Z}^\star_j$ is the Born effective charge tensor of the $j^\text{th}$ ion, $\vect{\varepsilon}_\infty$ is the high-frequency dielectric tensor that captures the electronic contribution to in-medium screening, $A_j$ is the atomic mass number, and $F_{N_j}(q) = \frac{3\,j_1(qr_j)}{qr_j}\,e^{-(qs)^2/2}$ (with $r_j= 1.14\, A_j^{1/3} \,\text{fm}$, $s = 0.9 \,\text{fm}$) is the Helm nuclear form factor \cite{Helm:1956zz} (which is close to unity for the DM masses considered in this work).

These benchmark models have highly complementary features. In a polar crystal, the dark photon couples to the Born effective charges of the ions, which have opposite signs within the primitive cell, and therefore dominantly induces out-of-phase oscillations corresponding to gapped optical phonon modes in the long-wavelength limit. By contrast, the hadrophilic scalar mediator couples to all ions with the same sign, and therefore dominantly excites gapless acoustic phonons that correspond to in-phase oscillations in the long-wavelength limit. There is also a difference between a light and heavy mediator due to the mediator form factor in Eq.~\eqref{eq:Fmed}. Noting that $\vect{Y}_j$ scales with $q$ and the energy conserving delta function contributes a factor of $q^{-1}$ (see Eq.~\eqref{eq:g_fun} below), we see that for a heavy mediator, the integral scales as $\int dq\, q^3\omega^{-1}$ and so is always dominated by large $q$. For a light mediator, on the other hand, the integral scales as $\int dq\, q^{-1}\omega^{-1}$. So for optical phonons with $\omega\sim q^0$, it receives similar contributions from all $q$, whereas for acoustic phonons, it is dominated by small $q$ where $\omega\sim q$.

\subsection{Daily Modulation}
\label{subsec:daily_mod_factors}

In the rate formula Eq.~\eqref{eq:rate}, the time dependence comes from the DM's velocity distribution $f_\chi(\vect{v},t)$, specifically via the Earth's velocity $\vve(t)$ that boosts the distribution. Concretely, we take
\begin{equation}
f_\chi(\vect{v},t) = \tfrac{1}{N_0} \exp\Bigl[-\tfrac{(\vect{v}+\vve(t))^2}{v_0^2}\Bigr]\, \Theta\bigl(\vesc-|\vect{v}+\vve(t)|\bigr)\,,
\label{eq:f_chi}
\end{equation}
where $N_0$ is a normalization constant such that $\int d^3v\, f_\chi(\vect{v}) = 1$, $\vesc = 600$ km/s is the galactic escape velocity, and $v_0=230$\,km/s as mentioned above. Assuming the detector is fixed on the Earth, in the lab frame $\vve$ becomes a function of time that is approximately periodic over a sidereal day as a result of the Earth's rotation. As a default setup, we adopt the detector orientation in Refs.~\cite{Griffin:2018bjn, Coskuner:2019odd, Trickle:2019nya}, for which, independent of the detector's location,
\begin{equation}
\vve(t) = \ve
\begin{pmatrix}
\sin{\theta_\text{e}\sin\phi(t)}\\
\sin{\theta_\text{e}} \cos{\theta_\text{e}} \left( \cos{\phi(t)} - 1 \right) \\
\cos^2{\theta_\text{e}}+ \sin^2{\theta_\text{e}}\cos{\phi(t)}
\end{pmatrix}\,,
\label{eq:ve_default}
\end{equation}
where $\ve=240\,$km/s, $\theta=42^\circ$, and $\phi(t)=2\pi\bigl(\frac{t}{24\,\text{hr}}\bigr)$. It is this periodicity of the {\it direction} of $\vve(t)$ that induces the daily modulation in $R(t)$ we study in this work.\footnote{Annual modulation is also present, due to the change of the {\it magnitude} of $\vve$, as in any terrestrial direct detection experiment. Here we fix $\ve=240\,$km/s and focus on the daily modulation signal, which is unique to anisotropic (crystal) targets.}

\begin{figure*}[t]
	\includegraphics[width=0.6\textwidth]{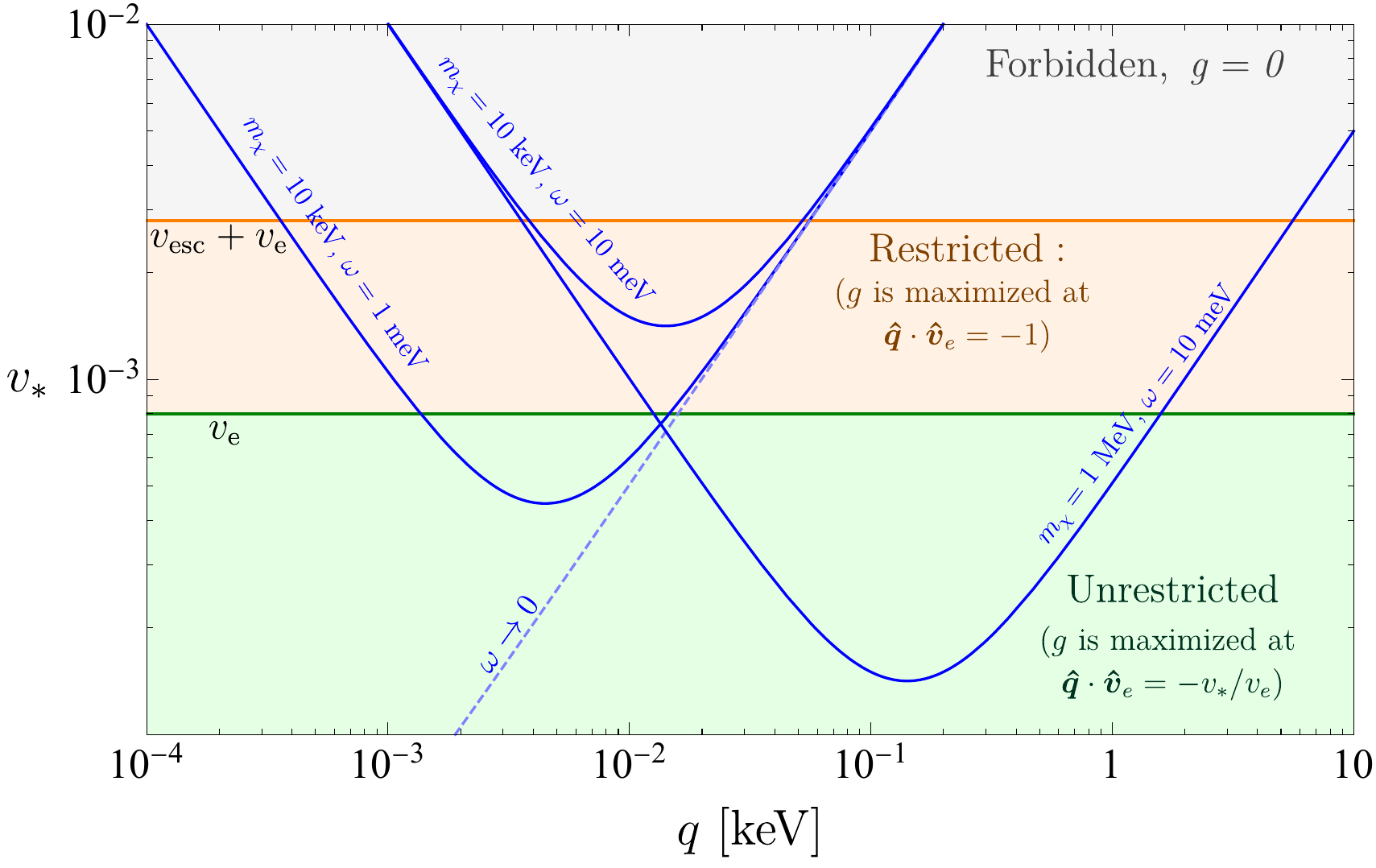}\\[10pt]
	\includegraphics[width=1.1\textwidth]{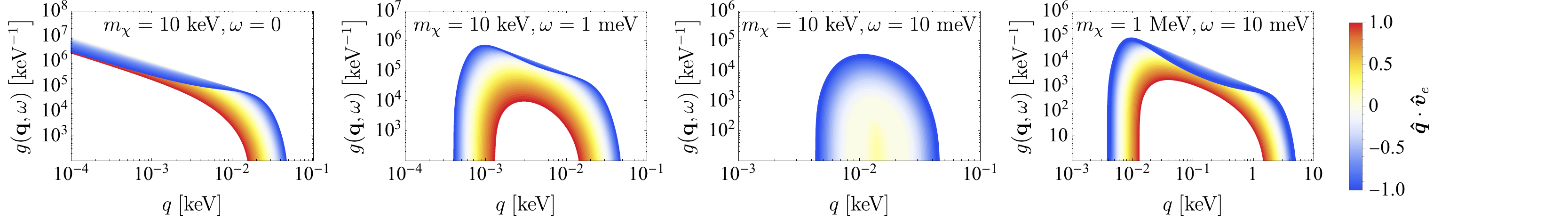}
	\caption{\textbf{Top:} To understand the kinematic function, $g(\vect{q},\omega)$, defined in Eq.~\eqref{eq:g_fun}, we plot $v_* \equiv \frac{q}{2m_\chi} + \frac{\omega}{q}$ as a function of $q$ (blue) for various $m_\chi$ and $\omega$ values. Comparing $v_*$ to $\ve$ and $\ve+\vesc$ we can qualitatively reconstruct the shape of $g(\vect{q},\omega)$, as discussed in the text.
		\textbf{Bottom:} $g(\vect{q},\omega)$ vs.\ $q$ for several fixed $m_\chi$, $\omega$ values, with varying $\vect{\hat{q}}\cdot\vveh$. The kinematic function weights different $\vect{\hat q}$ directions according to their angle with respect to $\vve(t)$, which ultimately leads to a daily modulating rate.}
	\label{fig:g_fun}
\end{figure*}

With the specific form of $f_\chi$ in Eq.~\eqref{eq:f_chi}, the velocity integral in Eq.~\eqref{eq:rate} can be done analytically~\cite{Griffin:2018bjn,Coskuner:2019odd,Trickle:2019nya}. We define
\begin{eqnarray}
g(\vect{q},\omega,t) &\equiv& \int d^3 v \, f_\chi(\vect{v},t) \,2\pi \delta(\omega-\omega_{\vect{q}}) \nonumber\\
&=& \tfrac{2\pi^2 v_0^2}{N_0 q} \Bigl\{\exp \Bigl[-\tfrac{(v_-(\vect{q},\omega,t))^2}{v_0^2}\Bigr] -\,\exp \Bigl[-\tfrac{\vesc^2}{v_0^2}\Bigr]\Bigr\}\,,\nonumber\\
\label{eq:g_fun}
\end{eqnarray}
where
\begin{equation}
v_-(\vect{q},\omega,t) = \min \Bigl( \Bigl|\vect{\hat q}\cdot\vve(t) +\tfrac{q}{2m_\chi} +\tfrac{\omega}{q}\Bigr|\,,\; \vesc \Bigr) \,.
\label{eq:v_minus}
\end{equation}
We will refer to $g(\vect{q},\omega,t)$ as the \textit{kinematic function}. 
The rate formula Eq.~\eqref{eq:rate} then becomes 
\begin{eqnarray}
R(t) = \frac{1}{\rho_T} \frac{\rho_\chi}{m_\chi} \frac{\pi\overline\sigma_{\psi}}{\mu_{\chi\psi}^2} \int&&\frac{d^3q}{(2\pi)^3} \,{\cal F}_\text{med}^2(q) \nonumber\\
&& \sum_\nu S'_\nu\bigl(\vect{q}) \,g(\vect{q},\omega_{\nu,\vect{k}},t) \,.
\label{eq:Rt_g}
\end{eqnarray}
With the rate written in this form, the time dependence now comes from the $v_-$ function contained in $g(\vect{q},\omega,t)$. As we will discuss in detail in the rest of this subsection, the origin of daily modulation is as follows. First, the kinematic function $g(\vect{q},\omega,t)$ selects a region of $\vect{q}$ space at each time of the day that is strongly correlated with $\vve(t)$. For anisotropic targets, this then results in a modulating rate after the $\vect{q}$ integral in Eq.~\eqref{eq:Rt_g}. Intuitively, the DM wind hits the target from different directions throughout the day, some of which may induce a stronger response than others.

\subsubsection{Kinematic Function}
\label{subsubsec:kinematics}

The kinematic function $g(\vect{q},\omega_{\nu,\vect{k}},t)$ can be viewed as a weight function: for each phonon branch $\nu$, the integrand in Eq.~\eqref{eq:Rt_g}, ${\cal F}_\text{med}^2(q) \,S'_\nu\bigl(\vect{q})$, is weighted toward momentum transfers $\vect{q}$ that maximize the $g$ function or, equivalently, minimize $v_-$ defined in Eq.~\eqref{eq:v_minus}. To visualize this minimization, we plot 
\begin{align}
    v_* \equiv \frac{q}{2m_\chi}+\frac{\omega}{q}
    \label{eq:v_star}
\end{align}
as a function of $q$ in the top panel of Fig.~\ref{fig:g_fun}. Setting $\omega$ to a constant approximates the case of optical phonons, which have relatively flat dispersions, whereas $\omega \to 0$ corresponds to the case of acoustic phonons, for which $\omega/q$ is bounded by the sound speed, which is typically much smaller than the DM's velocity. We can identify three distinct regions (as shown with different colors in the plot):
\begin{itemize}
	\item For $v_* \ge \vesc+\ve$, we have $v_-=\vesc$ and therefore $g=0$ for all $\vect{\hat q}$ directions. This is the kinematically forbidden region.
	\item For $\ve \le v_* < \vesc+\ve$, the $g$ function is maximized at $\vect{\hat q}\cdot\vveh=-1$.
    \item For $v_* \le \ve$, the $g$ function is nonzero for all $\vect{\hat q}$ directions, and is maximized at $\vect{\hat{q}} \cdot \vveh = -v_*/v_e$. In the large $m_\chi$, small $\omega$ limit, $v_* \rightarrow 0$, and therefore the $g$ function is maximized when $\vect{\hat{q}} \cdot \vveh = 0$.
\end{itemize}
These behaviors are seen in the lower panels of Fig.~\ref{fig:g_fun} (see also Ref.~\cite{Coskuner:2019odd}), where we plot $g(\vect{q},\omega)$ as a function of $q$ for fixed $m_\chi$, $\omega$, and with varying $\vect{\hat q}\cdot\vveh$. Note that in the $\omega\to0$ case, the $g$ function has support down to $q=0$, but the phase space integral for acoustic phonons is cut off at $q_\text{min} \simeq \frac{\omega_\text{min}}{c_s} = 2\times10^{-2}\,\text{keV}\,\bigl(\frac{\omega_\text{min}}{1\,\text{meV}}\bigr)\bigl(\frac{5\times10^{-5}}{c_s}\bigr)$, where $\omega_\text{min}$ is the detector's energy threshold, and $c_s$ is the sound speed (slope of the linear dispersion). From these plots we see that the kinematically favored region of $\vect{q}$ is strongly correlated with $\vveh(t)$ and, therefore, rotates with it throughout the day. This rotation then translates any target anisotropy into a detection rate that modulates daily.

\begin{figure*}[t]
	\centering
	\includegraphics[width=\textwidth]{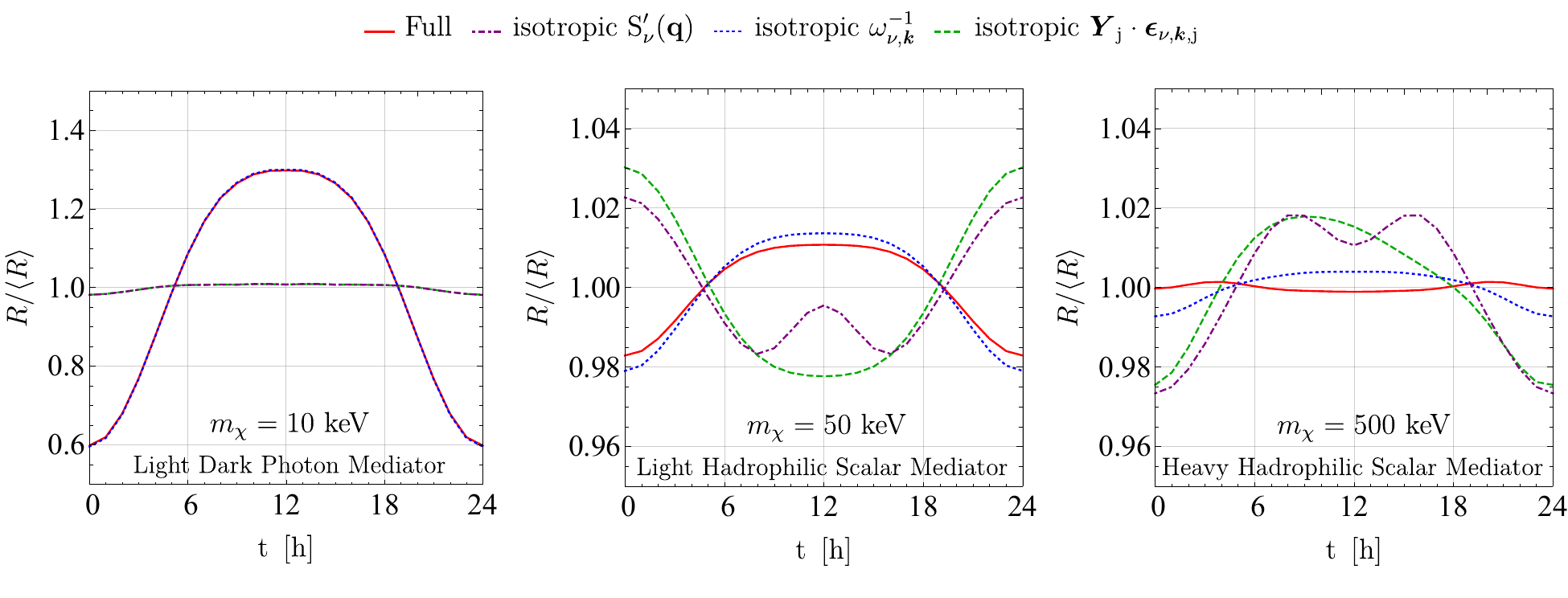}
	\caption{Comparison between the various sources of anisotropy a in SiO$_2$ target, for an example DM mass for each benchmark model. A 1\,meV energy threshold is assumed in all cases. As discussed in the text, anisotropy in the $\vect{Y}_j\cdot \vect{\epsilon}_{\nu, \vect{k}, j}$ factor in Eq.~\eqref{eq:structure_phonon} is the dominant factor in determining the daily modulation pattern.
	}
	\label{fig:SiO2-analysis-plot}
\end{figure*}

\subsubsection{Sources of Anisotropy}
\label{subsubsec:ani_target_response}

There are a number of possible sources of anisotropy, as we can infer from Eq.~\eqref{eq:Rt_g} and Eq.~\eqref{eq:structure_phonon}. First of all, the phonon energies $\omega_{\nu,\vect{k}}$ generically depend on the direction of $\vect{q}=\vect{k}+\vect{G}$. This means that the region selected by the kinematic function, as discussed above, does not preserve its shape as it rotates in $\vect{q}$ space. Also, the $\omega_{\nu,\vect{k}}^{-1}$ factor in Eq.~\eqref{eq:structure_phonon} is different in the dominating kinematic region at different times of the day, which adds to the daily modulation signal. 

The anisotropy in $\omega_{\nu,\vect{k}}$ has two contributing factors. First, phonon dispersions can be anisotropic as a result of crystal structures. For example, in h-BN, the sound speed of the longitudinal acoustic phonons differs by more than a factor of two between different $\vect{k}$ directions. Second, by the prescription explained above Eq.~\eqref{eq:S_sum_nu}, a sphere of constant $q$ outside the 1BZ does not map to a sphere of constant $k$ inside the 1BZ. Since the size of the 1BZ is typically ${\cal O}(\text{keV})$, and the DM velocity is ${\cal O}(10^{-3})$, this is relevant for $m_\chi \gtrsim $ MeV. Another related source of anisotropy is the $e^{i\vect{G}\cdot\vect{x}_j^0}$ factor in Eq.~\eqref{eq:structure_phonon}: a constant-$q$ sphere outside the 1BZ does not map onto a unique $\vect{G}$ vector. 

In addition to $\omega_{\nu,\vect{k}}$ and $e^{i\vect{G}\cdot\vect{x}_j^0}$ discussed above, the scalar product of the DM-ion coupling and phonon eigenvectors, $\vect{Y}_j\cdot \vect{\epsilon}_{\nu, \vect{k}, j}$, can also be anisotropic for a variety of reasons, depending on the DM model. For the hadrophilic scalar mediator model, $\vect{Y}_j\cdot \vect{\epsilon}_{\nu, \vect{k}, j}$ are simply proportional to the longitudinal components of phonon eigenvectors $\vect{\hat q}\cdot \vect{\epsilon}_{\nu, \vect{k}, j}$, so the anisotropy is determined by the extent to which the phonon eigenvectors deviate from transverse and longitudinal in different $\vect{\hat q}$ directions. For the dark photon mediator model, $\vect{Y}_j\cdot \vect{\epsilon}_{\nu, \vect{k}, j}$ are instead proportional to $\frac{\vect{\hat q}\cdot \vect{Z}^\star_j \cdot\vect{\epsilon}_{\nu, \vect{k}, j}}{\vect{\hat q}\cdot\vect{\varepsilon}_\infty\cdot\vect{\hat q}}$, so there are additional anisotropies if the Born effective charges $\vect{Z}^\star_j$ and dielectric tensor $\vect{\varepsilon}_\infty$ are not proportional to the identity. All these anisotropies are ultimately determined by the crystal structure.

We can carry out a simple exercise to see how the various sources of anisotropy discussed above contribute to the full daily modulation signal. As an example, we consider a \ce{SiO2} target, and pick one $m_\chi$ value for each benchmark model, as shown in the three panels of Fig.~\ref{fig:SiO2-analysis-plot}. We obtain the full rate normalized to its daily average, $R/\langle R\rangle$, as a function of time, as shown by the solid red curves labeled by ``full.'' We then artificially make the various factors in the rate formula isotropic and see how the modulation pattern changes. 

\begin{figure*}[t]
	\centering
	\includegraphics[width=\textwidth]{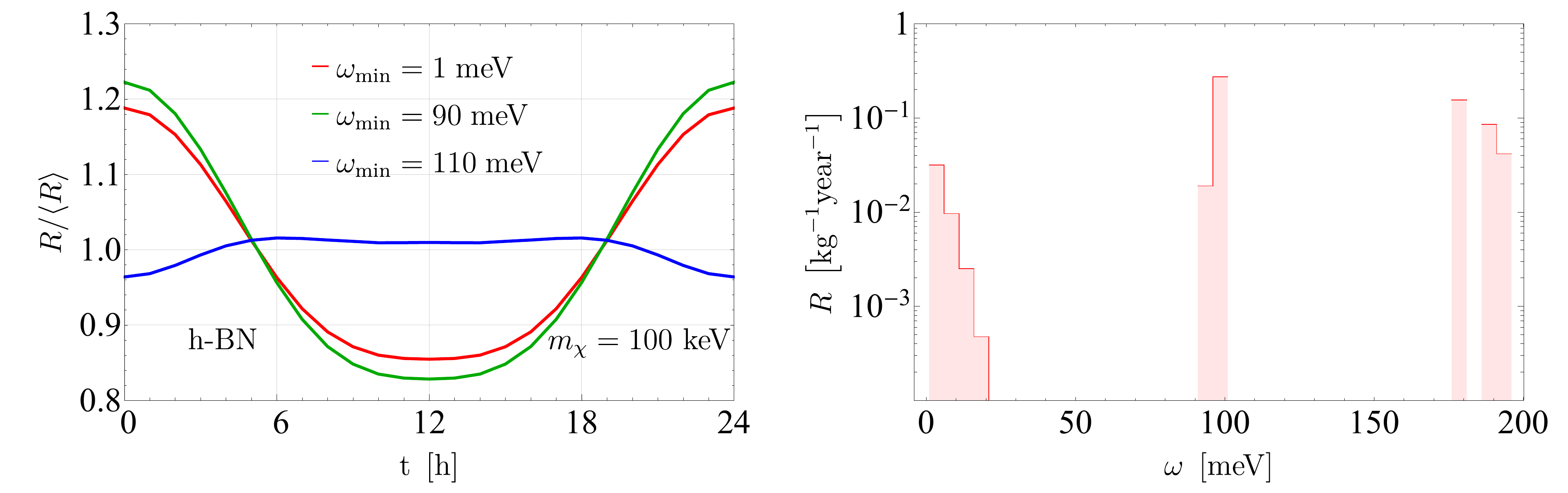}
	\caption{
		\textbf{Left:} Daily modulation for a h-BN target with various experimental thresholds, $\omega_{\rm{min}}$, assuming dark photon mediated scattering and $m_\chi = 100$ keV. 
		\textbf{Right:} Differential rate at $t=0$ for the same process assuming $\overline{\sigma}_e = 10^{-43} \; \text{cm}^2$. The daily modulation pattern is drastically different depending on whether the optical phonon modes just below 100 meV are included or excluded.\\[4pt]}
	\label{fig:hBN-combined-plot}
\end{figure*}

First, we make $S'_\nu(\vect{q})$ isotropic by setting $\omega_{\nu,\vect{k}}$ and $\vect{Y}_j\cdot \vect{\epsilon}_{\nu, \vect{k}, j}$ to their values at a specific direction ($\vect{\hat{q}}=\vect{\hat z}$), and setting $e^{i\vect{G}\cdot\vect{x}_j^0} \to 1$. This isolates the effect of the kinematic function $g(\vect{q}, \omega_{\nu,\vect{k}}, t)$ on daily modulation. The results are shown by the dot-dashed purple curves in Fig.~\ref{fig:SiO2-analysis-plot}, labeled ``isotropic $S'_\nu(\vect{q})$." In all three panels, we see that the ``isotropic $S'_\nu(\vect{q})$" curves are far from the full results (solid red curves), meaning that the anisotropy in $S'_\nu(\vect{q})$ plays an important role in determining the total modulation pattern. We find the same conclusion for the other materials and for other $m_\chi$, $\omega_\text{min}$ values.

\begin{figure*}[t]
	\includegraphics[width=0.5\textwidth]{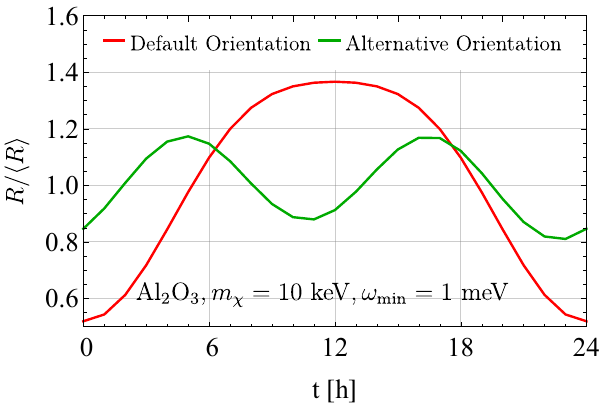}
    \caption{Effect of the crystal target orientation on the daily modulation pattern, for a sapphire target and the light dark photon mediator model as an example. The default orientation is the one adopted in Refs.~\cite{Griffin:2018bjn, Coskuner:2019odd, Trickle:2019nya} for which $\vve(t)$ is given by Eq.~\eqref{eq:ve_default}, and the alternative orientation is achieved by rotating the crystal $z$ axis by 60$^{\circ}$ clockwise around $\vect{\hat{n}} = (\vect{\hat{x}} + \vect{\hat{ y }} + \vect{\hat{z}})/\sqrt{3}$ (or equivalently, a $-60^{\circ}$ right-handed rotation around $\vect{\hat{n}}$.)\\[4pt]}
	\label{fig:GaAs-orientation}
\end{figure*}

\begin{figure*}[t]
	\centering
	\includegraphics[width=\textwidth]{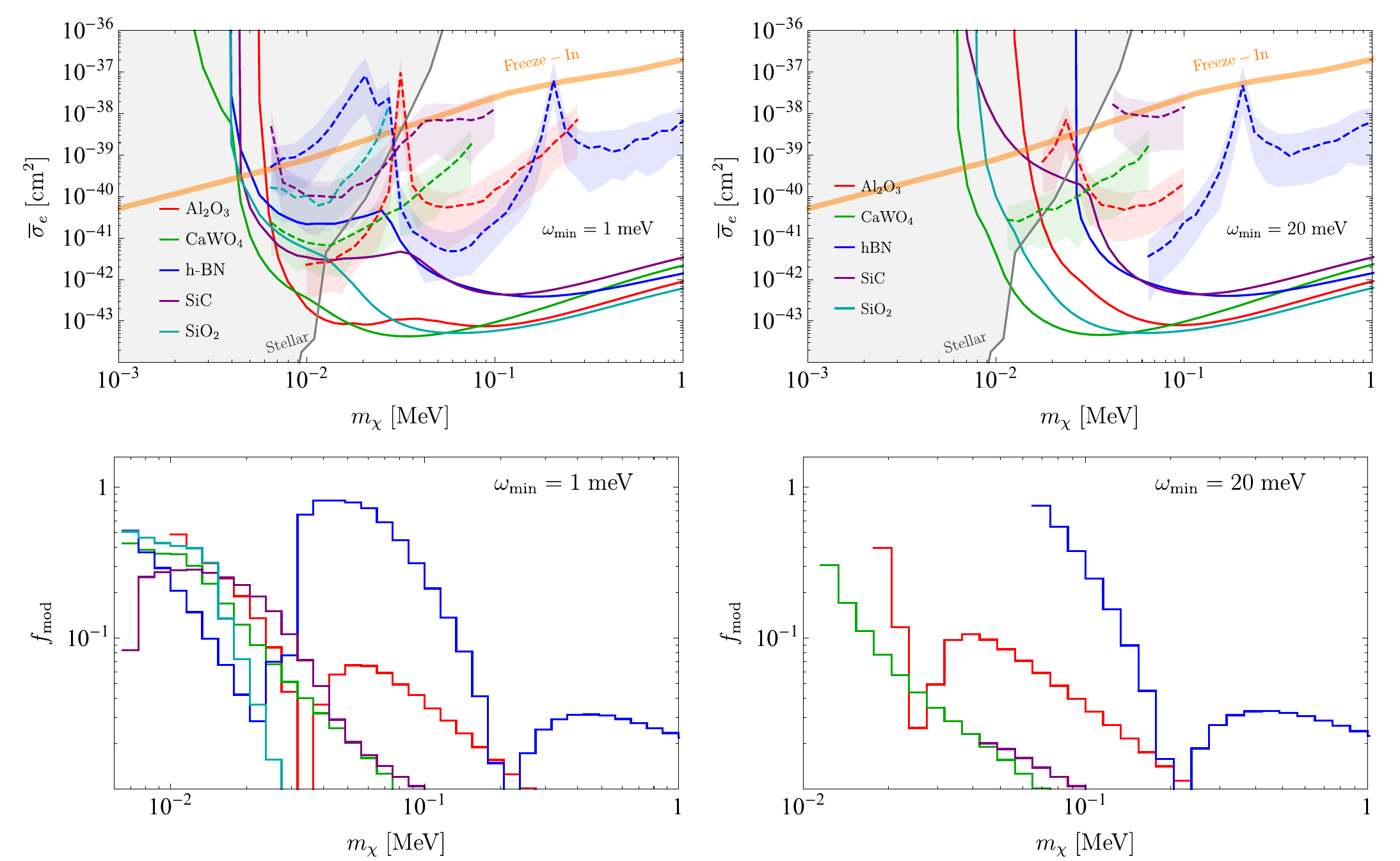}
    \caption{\textbf{Top:} Projected reach for the dark photon mediator model assuming 1\,meV and 20 meV detector energy thresholds and one kg-year exposure. Solid curves show the 95\% confidence level (CL) exclusion limits in the case of zero observed events, assuming no background. Dashed curves and the associated $\pm1\sigma$ bands show the modulation reach for DM masses with more than 1\% daily modulation, \ie\ cross sections for which we can reject the non-modulating hypothesis and establish the statistical significance of a modulating signal, as explained in App.~\ref{app:daily_mod_reach}. \textbf{Bottom:} Daily modulation amplitudes $f_\text{mod}$, defined in Eq.~\eqref{eq:f_mod}, for the same energy thresholds. Results are shown only for $m_\chi$ values where a material has substantial reach and $f_\text{mod}>10^{-2}$. The exact DM mass corresponding to a specific bar can be read off from the left edge of that bar.}
	\label{fig:dark-photon}
\end{figure*}

We can further dissect the anisotropy in $S'_\nu(\vect{q})$ by computing the daily modulation with $\omega_{\nu,\vect{k}}^{-1}$ or $\vect{Y}_j\cdot\vect{\epsilon}_{\nu,\vect{k},j}$ made isotropic by the same prescription as above; these are labeled ``isotropic $\omega_{\nu, \vect{k}}^{-1}$" (dotted blue curves) and ``isotropic $\vect{Y}_j \cdot \vect{\epsilon}_{\nu, \vect{k}, j}$" (dashed green curves) in Fig.~\ref{fig:SiO2-analysis-plot}, respectively. We see that the anisotropy in the $\vect{Y}_j\cdot\vect{\epsilon}_{\nu,\vect{k},j}$ factor contributes the most to daily modulation, as making it isotropic leads to the most significant deviations from the full results. We find the same is true for other materials.

We have also examined the effect of setting $e^{i\vect{G}\cdot\vect{x}_j^0}\to1$ in $S'_\nu(\vect{q})$ while leaving both $\omega_{\nu,\vect{k}}^{-1}$ and $\vect{Y}_j\cdot\vect{\epsilon}_{\nu,\vect{k},j}$ intact. This has a visible impact only when the region of $\vect{q}$ space just outside the 1BZ has a significant contribution to the rate; as $\vect{q}$ moves farther away from the 1BZ, summing over contributions from many different $\vect{G}$ vectors mitigates the effect. For the dark photon mediator model, this explains the enhanced daily modulation at $m_\chi\gtrsim$ MeV (see Fig.~\ref{fig:dark-photon} below). For the light hadrophilic scalar mediator model, there is no significant effect since the $\vect{q}$ integral is dominated by small $q$. For the heavy hadrophilic scalar mediator model, in contrast, the $\vect{q}$ integral is dominated by large $q$, so the enhancement happens in a window around $m_\chi\sim$ MeV (see Fig.~\ref{fig:massive-hadrophilic} below).

\subsubsection{Effects of Experimental Setup}
\label{subsubsec:experimental_effects}

The daily modulation pattern can also be significantly affected by experimental factors, including in particular the detector's energy threshold and the orientation of the target crystal. The energy threshold $\omega_\text{min}$ can be important if phonon modes at different energies have different modulation patterns. As an example, we show in the left panel of Fig.~\ref{fig:hBN-combined-plot} the daily modulation in h-BN for several different values of $\omega_\text{min}$, for the dark photon mediator model with $m_\chi=100$\,keV. The distinct daily modulation curves can be understood from the differential rate plot in the right panel of Fig.~\ref{fig:hBN-combined-plot}. We see that the phonon modes just below 100\,meV dominate the total rate, so long as $\omega_\text{min}$ is below this, and they drive the daily modulation pattern. On the other hand, if $\omega_\text{min}>100$\,meV, these modes are no longer accessible, and the daily modulation is instead induced by phonon modes at energies higher than about 175\,meV, for which the rate has a very different time dependence.

Meanwhile, the orientation of the crystal determines the function $\vve(t)$, and hence the daily modulation pattern. As an example, Fig.~\ref{fig:GaAs-orientation} compares the daily modulation patterns between our default setup, given in Eq.~\eqref{eq:ve_default}, and an (arbitrarily chosen) alternative orientation where the crystal $z$ axis is rotated by 60$^{\circ}$ clockwise around $\vect{\hat{n}} = (\vect{\hat{x}} + \vect{\hat{ y }} + \vect{\hat{z}})/\sqrt{3}$ (or equivalently, a $-60^{\circ}$ right-handed rotation around $\vect{\hat{n}}$.)

\section{Target Comparison}
\label{sec:target_comparison}

\begin{figure*}[t]
	\centering
	\includegraphics[width=\textwidth]{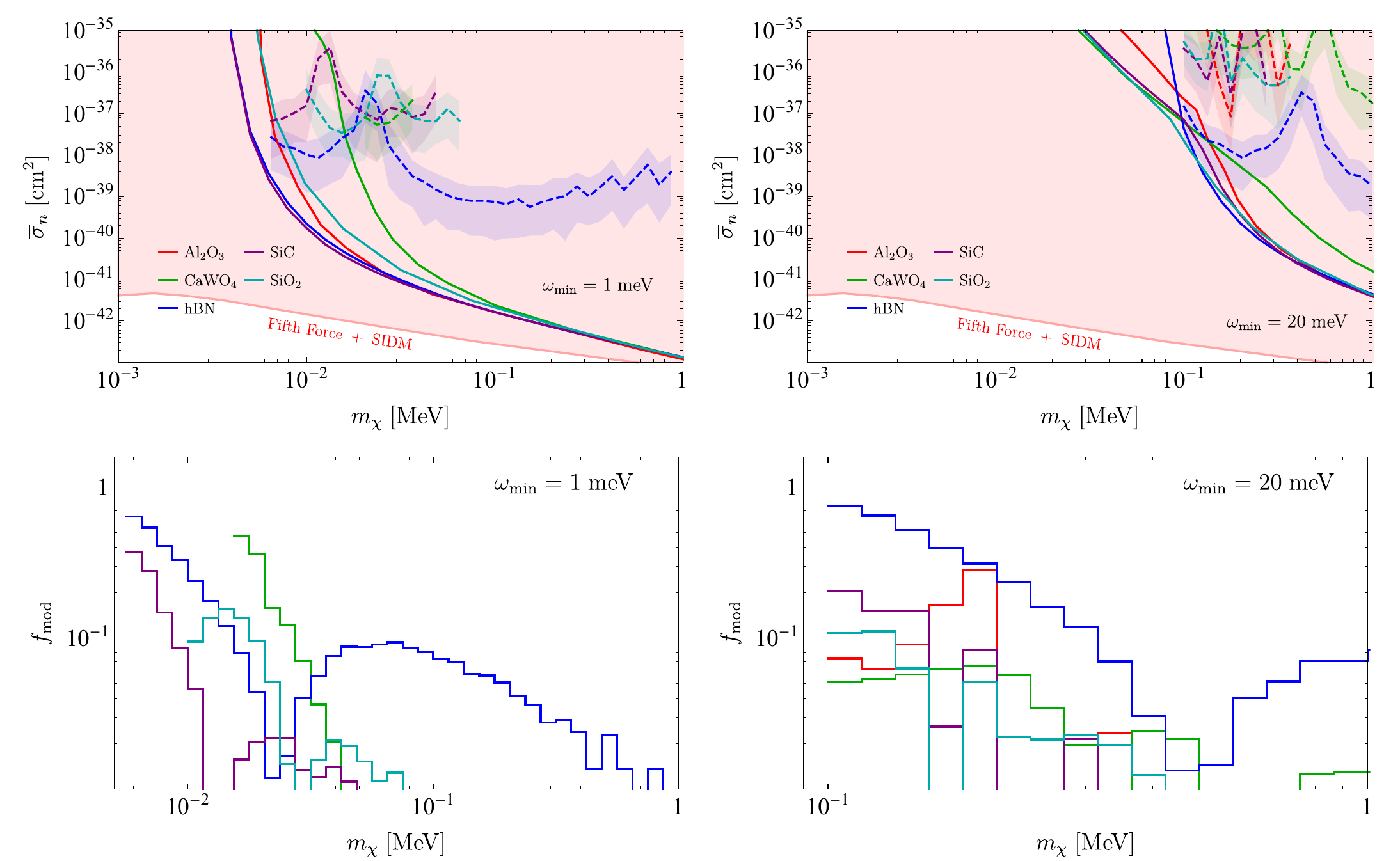}
	\caption{Same as Fig.~\ref{fig:dark-photon}, for the light hadrophilic scalar mediator model.}
	\label{fig:massless-hadrophilic}
\end{figure*}

\begin{figure*}[t]
	\centering
	
	\includegraphics[width=\textwidth]{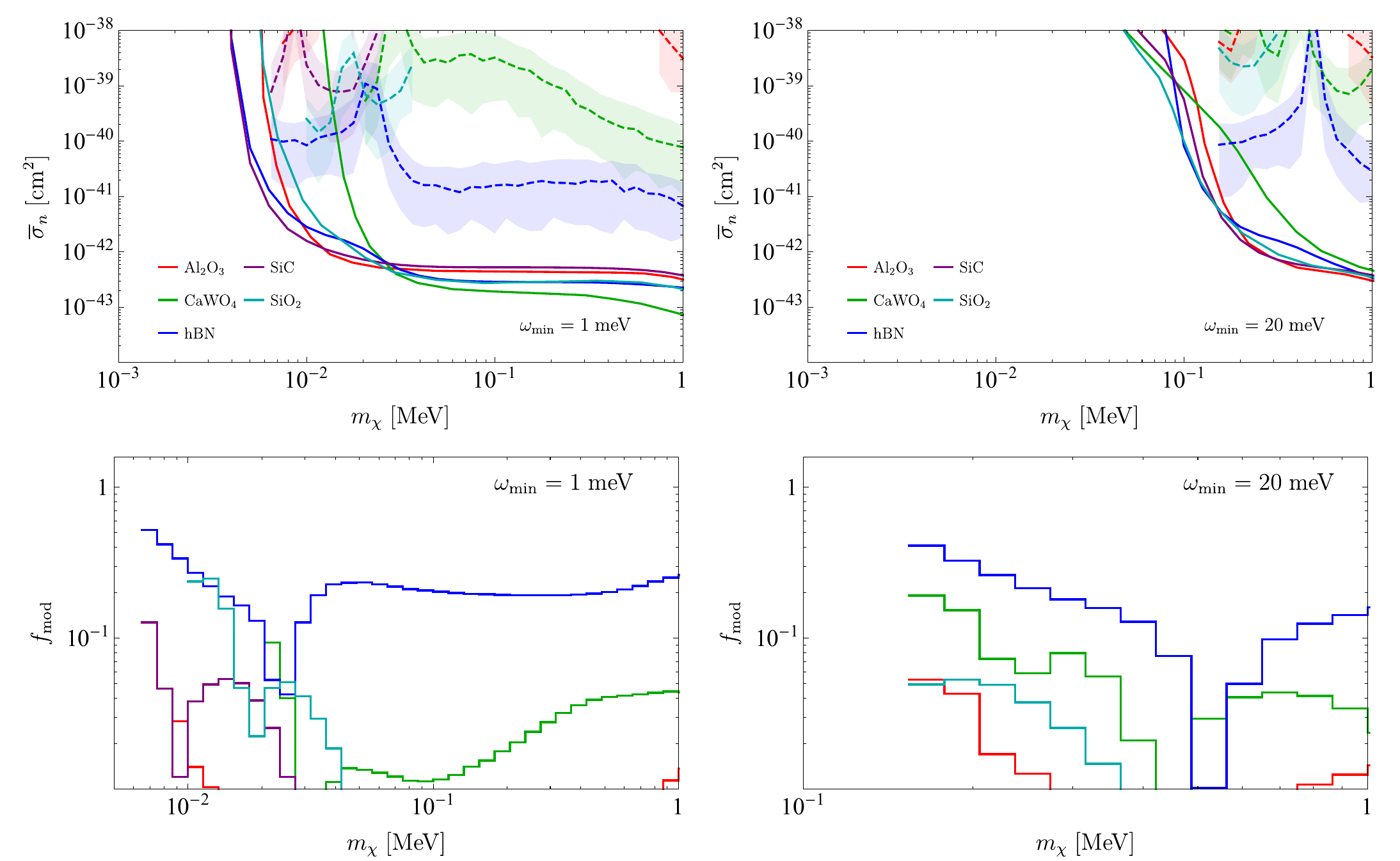}
	\caption{Same as Fig.~\ref{fig:dark-photon}, for the heavy hadrophilic scalar mediator model.}
	\label{fig:massive-hadrophilic}
\end{figure*}

\begin{figure*}[t]
	\centering
	\includegraphics[width=\textwidth]{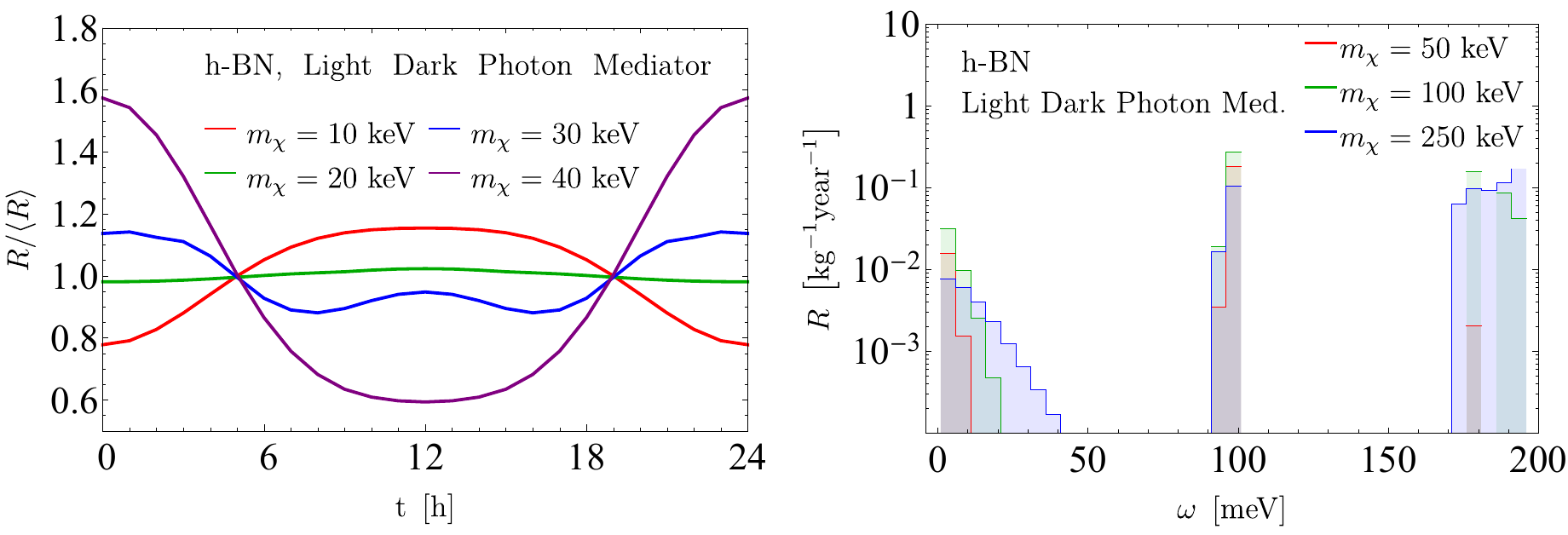}
	\caption{\textbf{Left:} Daily modulation for an h-BN target with various DM masses, assuming dark photon mediated scattering and $\omega_{\rm{min}}=1$\,meV. The change in modulation pattern is a result of the kinematically favored $\vect{\hat q}\cdot\vveh$ increasing from $-1$ toward 0 as $m_\chi$ increases. During the transition between different modulation patterns, an intermediate mass value around 20\,keV features a reduced modulation amplitude, which explains the peak in the modulation reach curve in the top-left panel of Fig.~\ref{fig:dark-photon}. A similar effect is also observed for the hadrophilic scalar mediator models in Figs.~\ref{fig:massless-hadrophilic} and \ref{fig:massive-hadrophilic}. \textbf{Right:} Differential rates at $t=0$ for several higher $m_\chi$ assuming $\overline{\sigma}_e = 10^{-43} \; \text{cm}^2$. Another transition between modulation patterns occurs when new phonon modes become dominant as $m_\chi$ increases, resulting in a second reduced modulation mass point, around 200\,keV, in  Fig.~\ref{fig:dark-photon}.}
	\label{fig:modreach-analysis}
\end{figure*}

Having discussed the physics underlying daily modulation, we now consider concrete target materials. Among the 26 materials studied~\cite{demo}, 19 are observed to have more than 1\% daily modulation for some DM masses in at least one of the benchmark models considered. In this section, we focus on the following five which are observed to have the highest daily modulation amplitudes: \ce{Al2O3}, \ce{SiO2}, \ce{SiC}, \ce{CaWO4} and h-BN. Among them, \ce{Al2O3}, \ce{SiO2} and \ce{SiC} have been proposed and recommended for near-future phonon-based experiments, while \ce{Al2O3} and \ce{CaWO4} are in use in the CRESST experiment. Meanwhile, h-BN is a highly anisotropic target with layered crystal structure that we have found to have exceptionally large daily modulation; while its experimental prospects have not been assessed, it serves as a useful benchmark for our theoretical study. We supplement this analysis with the remaining 14 materials with more than 1\% daily modulation (\ce{AlN}, \ce{CaF2}, \ce{GaN}, \ce{GaSb}, \ce{InSb}, \ce{LiF}, \ce{MgF2}, \ce{MgO}, \ce{NaF}, \ce{PbS}, \ce{PbSe}, \ce{PbTe}, \ce{ZnO}, \ce{ZnS}) in App.~\ref{app:daily_modulation_amplitudes}.

Our main results are shown in Figs.~\ref{fig:dark-photon}, \ref{fig:massless-hadrophilic} and \ref{fig:massive-hadrophilic}, for the dark photon mediator model and the light and heavy hadrophilic scalar mediator models, respectively. In the top panels of each figure, we show both the projected exclusion limits (solid) and the cross sections needed to distinguish the modulating signal and a non-modulating hypothesis in the event of discovery (dashed and shaded $\pm1\sigma$ bands), assuming 1 and 20\,meV detector energy thresholds. These energy thresholds have been envisioned with near-future advances in detector technology, and the primary motivation for these specific values is to differentiate the effects of acoustic and optical phonon dominated scattering. For the solid curves, we set $t=0$ when computing the rates for concreteness, and assume 3 events per kilogram-year exposure, corresponding to 95\,\% confidence level (CL) exclusion in a background-free experiment. The results for \ce{Al2O3}, \ce{CaWO4} and \ce{SiO2} were computed previously in Ref.~\cite{Griffin:2019mvc} (numerical errors in some of the materials in early versions of that reference have been corrected here and on the interactive webpage~\cite{demo}), and here we perform the calculation also for SiC and h-BN. For the dashed curves and the shaded bands for the modulation reach, we compute the number of events needed to reject the constant rate hypothesis at the 95\,\% confidence level by a prescription discussed in App.~\ref{app:daily_mod_reach}; they are truncated where the daily modulation falls below 1\%. 

In the lower panels of Figs.~\ref{fig:dark-photon}, \ref{fig:massless-hadrophilic} and \ref{fig:massive-hadrophilic}, we quantify the amount of daily modulation for several representative DM masses by
\begin{equation}
f_\text{mod} \equiv \frac{\max\bigl(|R-\langle R\rangle|\bigr)}{\langle R\rangle} \,,
\label{eq:f_mod}
\end{equation}
which characterizes the maximum deviation of detection rate throughout the day from the daily average $\langle R \rangle$. We shall refer to $f_\text{mod}$ as the daily modulation amplitude. The $f_\text{mod}$ plots give us an overview of the amount of daily modulation to expect. More detailed information on the daily modulation signal can be gained by plotting $R(t)/\langle R \rangle$, as in Figs.~\ref{fig:SiO2-analysis-plot}, \ref{fig:hBN-combined-plot} and \ref{fig:GaAs-orientation}, for each DM mass and energy threshold; we provide these plots on the interactive webpage~\cite{demo}.

We have considered detector energy thresholds $\omega_\text{min}=1$\,meV and 20\,meV. For the dark photon mediator model (Fig.~\ref{fig:dark-photon}), the energy threshold does not have a significant impact on either the reach or the daily modulation amplitude, except at the lowest $m_\chi$ values. This is because gapped optical phonons dominate the rate as long as they are above $\omega_\text{min}$ and the DM is heavy enough to excite them. For the hadrophilic scalar mediator models (Figs.~\ref{fig:massless-hadrophilic} and \ref{fig:massive-hadrophilic}), on the other hand, gapless acoustic phonons dominate and, as a result, both the reach and the daily modulation amplitude are sensitive to $\omega_\text{min}$. Generally, a higher energy threshold tends to amplify the daily modulation since the kinematically accessible phase space becomes limited, as discussed in detail in Sec.~\ref{subsubsec:kinematics}. Similarly, the daily modulation amplitude tends to increase at the lowest $m_\chi$ considered because of phase space restrictions. The enhanced daily modulation in these cases comes at the price of a lower total rate, so there is a trade-off between better overall sensitivity and a higher daily modulation signal. This is reflected by the dashed modulation reach curves in the top panels of each figure, which ascend at lower masses since the rate also vanishes.

From Figs.~\ref{fig:dark-photon}, \ref{fig:massless-hadrophilic} and \ref{fig:massive-hadrophilic}, we see that h-BN consistently outperforms all other materials in terms of the daily modulation amplitude, which reaches ${\cal O}(1)$ for some $m_\chi$ and $\omega_\text{min}$ values. This is due to the layered crystal structure which means that the momentum transfers perpendicular and parallel to the layers lead to very different target responses. Among the other materials, \ce{Al2O3}, \ce{CaWO4} and \ce{SiC} are also competitive targets for the dark photon mediator model at $m_\chi\lesssim 100\,$keV, and \ce{CaWO4} shows percent level daily modulation across a wide range of DM masses for the heavy scalar mediator model.

It is also worth noting that the modulation reach curves and $f_\text{mod}$ often exhibit a nontrivial dependence on $m_\chi$. In particular, for given target material and $\omega_\text{min}$, there can be $m_\chi$ values where the modulation signal diminishes. For example, for dark photon mediated scattering, h-BN with $\omega_\text{min}=1\,$meV has two such low-$f_\text{mod}$ mass points at around $20\,$keV and 200\,keV, corresponding to the peaks of the modulation reach curve in the top-left panel of Fig.~\ref{fig:dark-photon}.

Generally, low-$f_\text{mod}$ points at low $m_\chi$ result from the change in $\vect{\hat q}\cdot\vveh$ favored by the kinematic function. As discussed in Sec.~\ref{subsubsec:kinematics}, as $m_\chi$ increases, the favored $\vect{\hat q}\cdot\vveh$ increases from $-1$ toward 0. As $\vveh$ changes with time (\eg\ as in Eq.~\eqref{eq:ve_default}), a given $\vect{\hat q}\cdot\vveh$ probes the crystal's $S'_\nu(\vect{q})$ along a set of $\vect{\hat q}$ directions that modulates, and the modulation pattern depends on the kinematically favored $\vect{\hat q}\cdot\vveh$ value. We verify this expectation in the left panel of Fig.~\ref{fig:modreach-analysis} for h-BN. In this case, the modulation pattern flips as $m_\chi$ increases from 10\,keV to 40\,keV, and an approximate cancelation occurs around 20\,keV. Note, however, that the daily modulation sensitivity may be recovered by analyzing the differential rates $\frac{dR(t)}{d\omega}$.

The low-$f_\text{mod}$ points at higher $m_\chi$, on the other hand, are explained by new phonon modes with different modulation patterns becoming kinematically accessible as $m_\chi$ increases. Again focusing on h-BN as an example, we see from the right panel of Fig.~\ref{fig:modreach-analysis} that while the dominant phonon modes are the $\sim 100$\,meV modes for $m_\chi=50$\,keV and 100\,keV, the modes above 150 meV take over as $m_\chi$ increases to 250\,keV. The reduced modulation sensitivity at $m_\chi\simeq200\,$keV results from the transition between the two regimes.

\section{Conclusions}
\label{sec:conclusions}

As new experiments focused on light DM detection with single optical and acoustic phonons begin an R\&D phase~\cite{tesseract}, it is important and timely to understand which target crystals have the optimal sensitivity to well-motivated DM models. This includes not only the sensitivity to the smallest interaction cross section for a given DM model, but also the ability to extract a smoking gun signature for DM that can be distinguished from background. Daily modulation provides such a unique fingerprint. In this work, we have carried out a comparative study of daily modulation signals for several benchmark models, where DM scattering is mediated by a dark photon or hadrophilic scalar mediator. Our results supplement the information on the cross section reach obtained previously in Ref.~\cite{Griffin:2019mvc}, and provide further theoretical guidance to the optimization of near future phonon-based experiments.

Based on our analysis of 26 crystals, we observe that there is often a trade-off between detection rate, modulation amplitude, and experimental feasibility. For example, for dark photon mediated scattering, \ce{Al2O3} (sapphire), \ce{CaWO4} and \ce{SiO2} ($\alpha$-quartz) outperform h-BN in terms of their sensitivities to the total rate; h-BN's daily modulation signal, however, is significantly stronger. Still, despite having the largest daily modulation amplitude, h-BN will likely be difficult to fabricate as a large ultra-pure single crystal target. Overall, \ce{Al2O3} and \ce{CaWO4} provide perhaps the optimal balance between the overall reach and the daily modulation signal, and have both already been used in direct detection experiments.

Beyond the results presented in this paper, we also publish an interactive webpage~\cite{demo}, where additional results can be generated from our calculations of single phonon excitation rates and their daily modulation.

\vspace{12pt}
{\em Acknowledgments.}
We thank Sin\'ead Griffin and Katie Inzani for previous collaboration on DFT calculations (published in Ref.~\cite{Griffin:2019mvc}) utilized in this work. We also thank Matt Pyle for useful discussions. 
This material is based upon work supported by the U.S.\ Department of Energy, Office of Science, Office of High Energy Physics, under Award Number DE-SC0021431, and the Quantum Information Science Enabled Discovery (QuantISED) for High Energy Physics (KA2401032).

\vspace{12pt}
\appendix

\begin{figure*}
	\centering
	\includegraphics[width=0.5\linewidth]{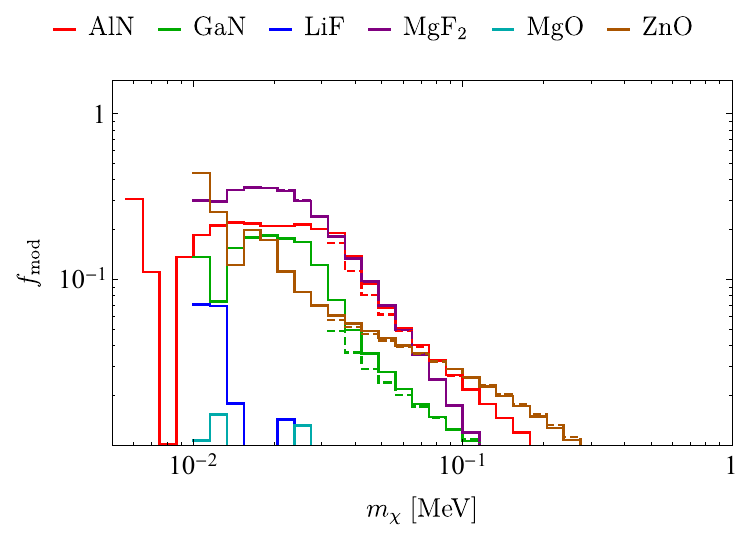}
	\caption{Daily modulation amplitudes for the dark photon mediator model. Solid and dashed curves assume energy thresholds of 1\,meV and 20\,meV, respectively. Among the materials studied, only those that have a modulation amplitude greater than 1\% for at least one $m_\chi$ value (at which the material has substantial reach) are shown. As in the lower panels of Figs.~\ref{fig:dark-photon}, \ref{fig:massless-hadrophilic} and \ref{fig:massive-hadrophilic} in the main text, the low mass values where the rate diminishes are excluded for each material. Therefore the shown modulation amplitudes correspond to the mass values where the materials have reach.}
	\label{fig:amplitude-darkphoton-allmaterials}
\end{figure*}

\begin{figure*}
	\centering
	\includegraphics[width=\linewidth]{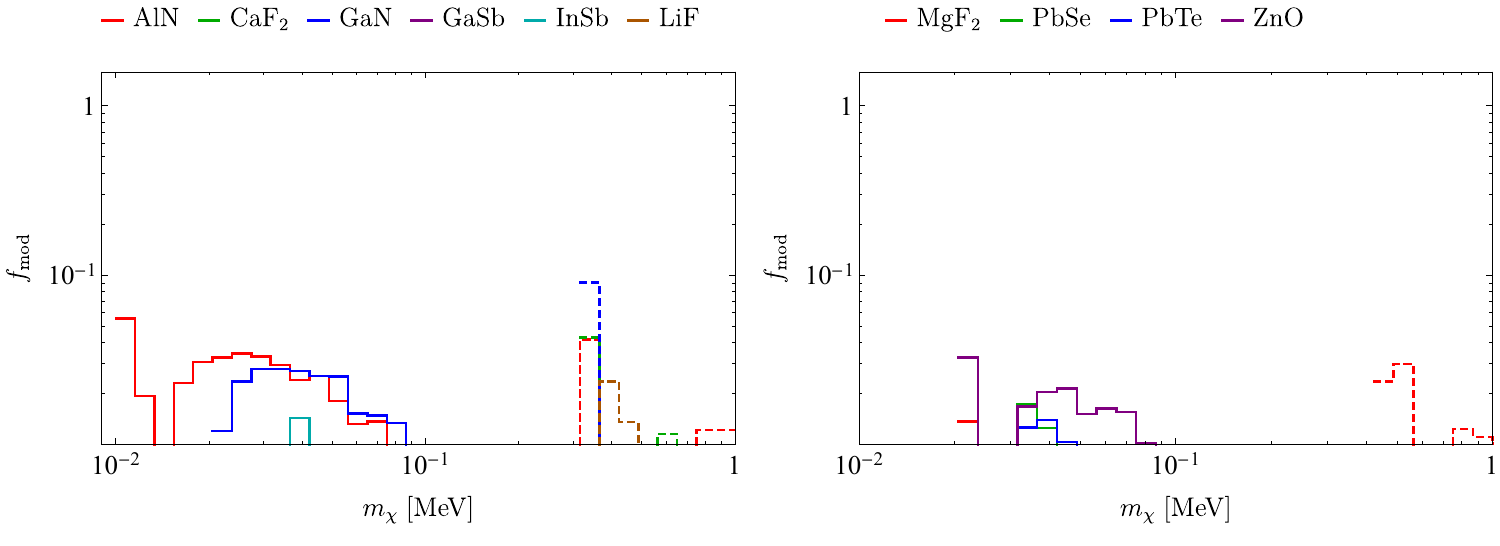}
	\caption{
		Same as Fig.~\ref{fig:amplitude-darkphoton-allmaterials}, for the light hadrophilic scalar mediator model.
	}
	\label{fig:amplitude-hadrophilic-massless-allmaterials}
\end{figure*}

\begin{figure*}
	\centering
	\includegraphics[width=\linewidth]{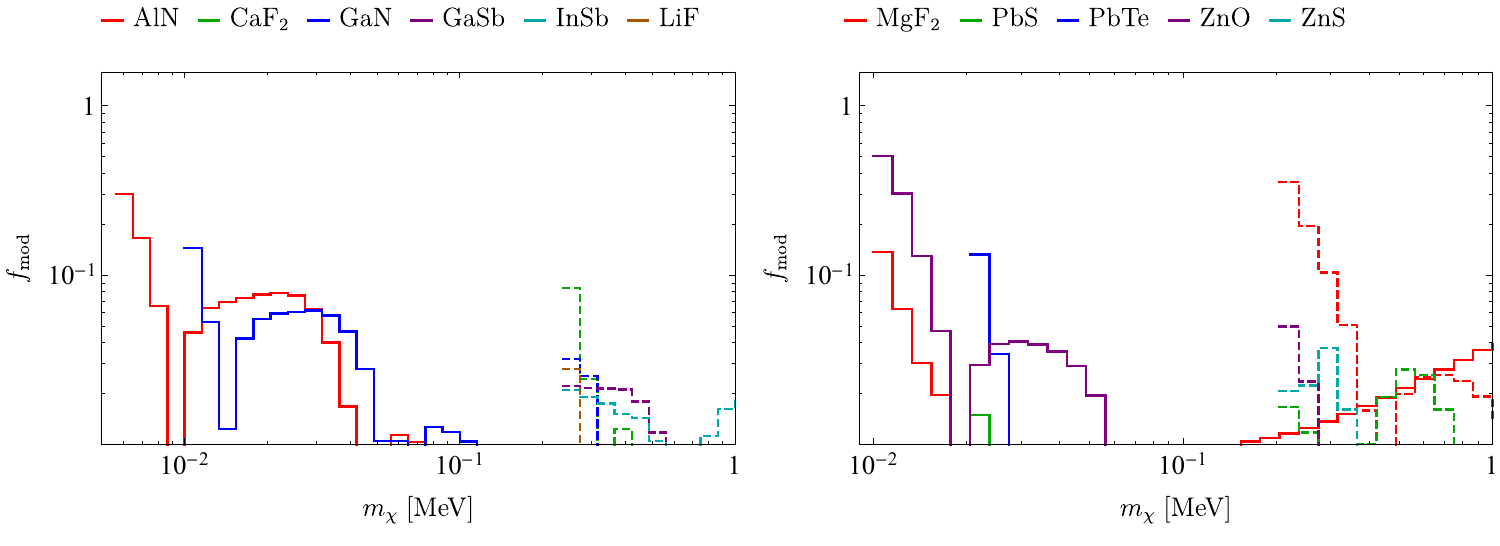}
	\caption{
		Same as Fig.~\ref{fig:amplitude-darkphoton-allmaterials}, for the heavy hadrophilic scalar mediator model.
	}
	\label{fig:amplitude-hadrophilic-massive-allmaterials}
\end{figure*}

\section{Daily Modulation Amplitudes for Additional Materials}
\label{app:daily_modulation_amplitudes}

In addition to the materials discussed in the main text, we have also investigated the daily modulation of the full list of materials considered in Ref.~\cite{Griffin:2019mvc}. Their projected reach curves were already computed in Ref.~\cite{Griffin:2019mvc} and are included on the interactive webpage~\cite{demo}. Here we only show the daily modulation amplitude $f_\text{mod}$, defined in Eq.~\eqref{eq:f_mod}, as in the lower panels of Figs.~\ref{fig:dark-photon}, \ref{fig:massless-hadrophilic} and \ref{fig:massive-hadrophilic} in the main text. The results for the dark photon mediator model, the light hadrophilic scalar mediator model and the heavy hadrophilic scalar mediator model are shown in Figs.~\ref{fig:amplitude-darkphoton-allmaterials}, \ref{fig:amplitude-hadrophilic-massless-allmaterials} and \ref{fig:amplitude-hadrophilic-massive-allmaterials}, respectively. Only materials with $f_\text{mod}\ge 10^{-2}$ for at least one $m_\chi$ value (at which the material has substantial reach) are shown in each case.

\section{Calculation of the Modulation Reach}
\label{app:daily_mod_reach}

To establish the statistical significance of a modulating signal, we find the expected number of events needed to reject the non-modulating hypothesis using the following procedure. 
For a given DM model, with the DM mass $m_\chi$ and experimental energy threshold $\omega_\text{min}$ specified, we first obtain the modulating signal shape $r(t) \equiv R(t)/\langle R\rangle$ as explained in the main text. 
We divide a sidereal day into $N_\text{bins}=24$ equal-size bins, and denote the bin boundaries by $t_k = (k/N_\text{bins})$ days; this binning is fine enough to capture diverse modulation patterns of the large set of materials studied here.  
Given an expected number of events $N_\text{exp}$, we simulate a DM signal sample and a non-modulating sample by generating events following a Poisson distribution in each bin, with mean $\langle  N_k\rangle_\text{sig} \equiv N_\text{exp}\int_{t_{k-1}}^{t_k} r(t)\, dt/\text{day}$ and $\langle  N_k\rangle_\text{non-mod} \equiv N_\text{exp}/N_\text{bins}$ for the $k$th bin, respectively. 
We define our test statistic to be the difference between the Pearson's $\chi^2$ values when fitting the simulated data to the non-modulating  vs.\ modulating signal shapes. 
Concretely, suppose the number of events in the $k$th bin is $N_k$. 
The test statistic is given by 
\begin{equation}
\text{TS} = \sum_k \frac{(N_k-\langle  N_k\rangle_\text{non-mod})^2}{\langle  N_k\rangle_\text{non-mod}} -\sum_k \frac{(N_k-\langle  N_k\rangle_\text{sig})^2}{\langle  N_k\rangle_\text{sig}} \,.
\end{equation}
Given $N_\text{exp}$, we simulate events according to the modulating (DM signal) and non-modulating hypotheses for $N_\text{sample}=10^4$ times each, and obtain the distribution of TS for the modulating and non-modulating samples. 
For the non-modulating sample, we compute the 95 percentile value TS$_\text{non-mod,\,95\%}$. 
For the modulating signal sample, we compute the mean TS$_\text{sig,\,mean}$ and the $(50\pm 34)$ percentiles TS$_\text{sig,\,$\pm1\sigma$}$. 
These numbers tell us to what extent we can reject the non-modulating hypothesis: TS$_\text{non-mod,\,95\%}<$ TS$_\text{sig,\,mean}$ means we can reject the non-modulating hypothesis at 95\% CL on average, while TS$_\text{non-mod,\,95\%}<$ TS$_\text{sig,\,$\pm1\sigma$}$ means we can reject the non-modulating hypothesis at 95\% CL given a $\pm1\sigma$ statistical fluctuation of the signal. 
Repeating the calculation for many values of $N_\text{exp}$, we obtain the interpolating functions TS$_\text{non-mod,\,95\%} (N_\text{exp})$, TS$_\text{sig,\,mean} (N_\text{exp})$ and TS$_\text{sig,\,$\pm1\sigma$} (N_\text{exp})$. 
These allow us to solve for the $N_\text{exp}$ needed for TS$_\text{non-mod,\,95\%}$ to drop below TS$_\text{sig,\,mean}$, and for it to go below TS$_\text{sig,\,$\pm1\sigma$}$.  These then translate into cross sections assuming 1\,kg-yr exposure, represented by the modulation reach curves in Figs.~\ref{fig:dark-photon}, \ref{fig:massless-hadrophilic} and \ref{fig:massive-hadrophilic}. 
Note that the procedure here largely follows that in Ref.~\cite{Griffin:2018bjn}, but we have adopted a different test statistic that we find simpler to compute and interpret. 
We have checked that using instead the test statistic in Ref.~\cite{Griffin:2018bjn} produces very similar results in most cases.


\bibliography{bibliography}

\begin{thebibliography}{70}%
\makeatletter
\providecommand \@ifxundefined [1]{%
 \@ifx{#1\undefined}
}%
\providecommand \@ifnum [1]{%
 \ifnum #1\expandafter \@firstoftwo
 \else \expandafter \@secondoftwo
 \fi
}%
\providecommand \@ifx [1]{%
 \ifx #1\expandafter \@firstoftwo
 \else \expandafter \@secondoftwo
 \fi
}%
\providecommand \natexlab [1]{#1}%
\providecommand \enquote  [1]{``#1''}%
\providecommand \bibnamefont  [1]{#1}%
\providecommand \bibfnamefont [1]{#1}%
\providecommand \citenamefont [1]{#1}%
\providecommand \href@noop [0]{\@secondoftwo}%
\providecommand \href [0]{\begingroup \@sanitize@url \@href}%
\providecommand \@href[1]{\@@startlink{#1}\@@href}%
\providecommand \@@href[1]{\endgroup#1\@@endlink}%
\providecommand \@sanitize@url [0]{\catcode `\\12\catcode `\$12\catcode
  `\&12\catcode `\#12\catcode `\^12\catcode `\_12\catcode `\%12\relax}%
\providecommand \@@startlink[1]{}%
\providecommand \@@endlink[0]{}%
\providecommand \url  [0]{\begingroup\@sanitize@url \@url }%
\providecommand \@url [1]{\endgroup\@href {#1}{\urlprefix }}%
\providecommand \urlprefix  [0]{URL }%
\providecommand \Eprint [0]{\href }%
\providecommand \doibase [0]{https://doi.org/}%
\providecommand \selectlanguage [0]{\@gobble}%
\providecommand \bibinfo  [0]{\@secondoftwo}%
\providecommand \bibfield  [0]{\@secondoftwo}%
\providecommand \translation [1]{[#1]}%
\providecommand \BibitemOpen [0]{}%
\providecommand \bibitemStop [0]{}%
\providecommand \bibitemNoStop [0]{.\EOS\space}%
\providecommand \EOS [0]{\spacefactor3000\relax}%
\providecommand \BibitemShut  [1]{\csname bibitem#1\endcsname}%
\let\auto@bib@innerbib\@empty
\bibitem [{\citenamefont {Amar\'e}\ \emph {et~al.}(2020)\citenamefont {Amar\'e}
  \emph {et~al.}}]{Amare:2019ncj}%
  \BibitemOpen
  \bibfield  {author} {\bibinfo {author} {\bibfnamefont {J.}~\bibnamefont
  {Amar\'e}} \emph {et~al.},\ }\href
  {https://doi.org/10.1088/1742-6596/1468/1/012014} {\bibfield  {journal}
  {\bibinfo  {journal} {J. Phys. Conf. Ser.}\ }\textbf {\bibinfo {volume}
  {1468}},\ \bibinfo {pages} {012014} (\bibinfo {year} {2020})},\ \Eprint
  {https://arxiv.org/abs/1910.13365} {arXiv:1910.13365 [astro-ph.IM]}
  \BibitemShut {NoStop}%
\bibitem [{\citenamefont {Pr{\"o}bst}\ \emph {et~al.}(2002)\citenamefont
  {Pr{\"o}bst} \emph {et~al.}}]{Probst:2002qb}%
  \BibitemOpen
  \bibfield  {author} {\bibinfo {author} {\bibfnamefont {F.}~\bibnamefont
  {Pr{\"o}bst}} \emph {et~al.},\ }\bibfield  {booktitle} {\emph {\bibinfo
  {booktitle} {{Proceedings, 7th International Workshop on Topics in
  Astroparticle and Underground Physics (TAUP 2001): Gran Sasso, Assergi,
  L'Aquila, Italy, September 8-12, 2001}}},\ }\href
  {https://doi.org/10.1016/S0920-5632(02)01453-6} {\bibfield  {journal}
  {\bibinfo  {journal} {Nucl. Phys. Proc. Suppl.}\ }\textbf {\bibinfo {volume}
  {110}},\ \bibinfo {pages} {67} (\bibinfo {year} {2002})}\BibitemShut
  {NoStop}%
\bibitem [{\citenamefont {Angloher}\ \emph {et~al.}(2019)\citenamefont
  {Angloher} \emph {et~al.}}]{Angloher:2018fcs}%
  \BibitemOpen
  \bibfield  {author} {\bibinfo {author} {\bibfnamefont {G.}~\bibnamefont
  {Angloher}} \emph {et~al.} (\bibinfo {collaboration} {CRESST}),\ }\href
  {https://doi.org/10.1140/epjc/s10052-018-6523-4} {\bibfield  {journal}
  {\bibinfo  {journal} {Eur. Phys. J. C}\ }\textbf {\bibinfo {volume} {79}},\
  \bibinfo {pages} {43} (\bibinfo {year} {2019})},\ \Eprint
  {https://arxiv.org/abs/1809.03753} {arXiv:1809.03753 [hep-ph]} \BibitemShut
  {NoStop}%
\bibitem [{\citenamefont {Kluck}\ \emph {et~al.}(2020)\citenamefont {Kluck}
  \emph {et~al.}}]{Kluck:2020bdm}%
  \BibitemOpen
  \bibfield  {author} {\bibinfo {author} {\bibfnamefont {H.}~\bibnamefont
  {Kluck}} \emph {et~al.} (\bibinfo {collaboration} {CRESST}),\ }\href
  {https://doi.org/10.1088/1742-6596/1468/1/012038} {\bibfield  {journal}
  {\bibinfo  {journal} {J. Phys. Conf. Ser.}\ }\textbf {\bibinfo {volume}
  {1468}},\ \bibinfo {pages} {012038} (\bibinfo {year} {2020})}\BibitemShut
  {NoStop}%
\bibitem [{\citenamefont {Bernabei}\ \emph {et~al.}(2020)\citenamefont
  {Bernabei} \emph {et~al.}}]{Bernabei:2020mon}%
  \BibitemOpen
  \bibfield  {author} {\bibinfo {author} {\bibfnamefont {R.}~\bibnamefont
  {Bernabei}} \emph {et~al.},\ }\href
  {https://doi.org/10.1016/j.ppnp.2020.103810} {\bibfield  {journal} {\bibinfo
  {journal} {Prog. Part. Nucl. Phys.}\ }\textbf {\bibinfo {volume} {114}},\
  \bibinfo {pages} {103810} (\bibinfo {year} {2020})}\BibitemShut {NoStop}%
\bibitem [{\citenamefont {Aguilar-Arevalo}\ \emph {et~al.}(2020)\citenamefont
  {Aguilar-Arevalo} \emph {et~al.}}]{Aguilar-Arevalo:2020oii}%
  \BibitemOpen
  \bibfield  {author} {\bibinfo {author} {\bibfnamefont {A.}~\bibnamefont
  {Aguilar-Arevalo}} \emph {et~al.} (\bibinfo {collaboration} {DAMIC}),\ }\href
  {https://doi.org/10.1103/PhysRevLett.125.241803} {\bibfield  {journal}
  {\bibinfo  {journal} {Phys. Rev. Lett.}\ }\textbf {\bibinfo {volume} {125}},\
  \bibinfo {pages} {241803} (\bibinfo {year} {2020})},\ \Eprint
  {https://arxiv.org/abs/2007.15622} {arXiv:2007.15622 [astro-ph.CO]}
  \BibitemShut {NoStop}%
\bibitem [{\citenamefont {Agnes}\ \emph {et~al.}(2018)\citenamefont {Agnes}
  \emph {et~al.}}]{Agnes:2018ves}%
  \BibitemOpen
  \bibfield  {author} {\bibinfo {author} {\bibfnamefont {P.}~\bibnamefont
  {Agnes}} \emph {et~al.} (\bibinfo {collaboration} {DarkSide}),\ }\href
  {https://doi.org/10.1103/PhysRevLett.121.081307} {\bibfield  {journal}
  {\bibinfo  {journal} {Phys. Rev. Lett.}\ }\textbf {\bibinfo {volume} {121}},\
  \bibinfo {pages} {081307} (\bibinfo {year} {2018})},\ \Eprint
  {https://arxiv.org/abs/1802.06994} {arXiv:1802.06994 [astro-ph.HE]}
  \BibitemShut {NoStop}%
\bibitem [{\citenamefont {Jo}(2017)}]{Jo:2016qql}%
  \BibitemOpen
  \bibfield  {author} {\bibinfo {author} {\bibfnamefont {J.~H.}\ \bibnamefont
  {Jo}} (\bibinfo {collaboration} {DM-Ice}),\ }\bibfield  {booktitle} {\emph
  {\bibinfo {booktitle} {{Proceedings, 38th International Conference on High
  Energy Physics (ICHEP 2016): Chicago, IL, USA, August 3-10, 2016}}},\ }\href
  {https://doi.org/10.22323/1.282.1223} {\bibfield  {journal} {\bibinfo
  {journal} {PoS}\ }\textbf {\bibinfo {volume} {ICHEP2016}},\ \bibinfo {pages}
  {1223} (\bibinfo {year} {2017})},\ \Eprint {https://arxiv.org/abs/1612.07426}
  {arXiv:1612.07426 [physics.ins-det]} \BibitemShut {NoStop}%
\bibitem [{\citenamefont {Kim}(2015)}]{Kim:2015prm}%
  \BibitemOpen
  \bibfield  {author} {\bibinfo {author} {\bibfnamefont {K.}~\bibnamefont
  {Kim}} (\bibinfo {collaboration} {KIMS}),\ }in\ \href@noop {} {\emph
  {\bibinfo {booktitle} {{Proceedings, Meeting of the APS Division of Particles
  and Fields (DPF 2015): Ann Arbor, Michigan, USA, 4-8 Aug 2015}}}}\ (\bibinfo
  {year} {2015})\ \Eprint {https://arxiv.org/abs/1511.00023} {arXiv:1511.00023
  [physics.ins-det]} \BibitemShut {NoStop}%
\bibitem [{\citenamefont {Akerib}\ \emph
  {et~al.}(2020{\natexlab{a}})\citenamefont {Akerib} \emph
  {et~al.}}]{Akerib:2019zrt}%
  \BibitemOpen
  \bibfield  {author} {\bibinfo {author} {\bibfnamefont {D.~S.}\ \bibnamefont
  {Akerib}} \emph {et~al.} (\bibinfo {collaboration} {LUX}),\ }\href
  {https://doi.org/10.1103/PhysRevD.101.042001} {\bibfield  {journal} {\bibinfo
   {journal} {Phys. Rev. D}\ }\textbf {\bibinfo {volume} {101}},\ \bibinfo
  {pages} {042001} (\bibinfo {year} {2020}{\natexlab{a}})},\ \Eprint
  {https://arxiv.org/abs/1907.06272} {arXiv:1907.06272 [astro-ph.CO]}
  \BibitemShut {NoStop}%
\bibitem [{\citenamefont {Akerib}\ \emph
  {et~al.}(2020{\natexlab{b}})\citenamefont {Akerib} \emph
  {et~al.}}]{Akerib:2020gzy}%
  \BibitemOpen
  \bibfield  {author} {\bibinfo {author} {\bibfnamefont {D.~S.}\ \bibnamefont
  {Akerib}} \emph {et~al.} (\bibinfo {collaboration} {LUX}),\ }\href@noop {} {\
   (\bibinfo {year} {2020}{\natexlab{b}})},\ \Eprint
  {https://arxiv.org/abs/2003.11141} {arXiv:2003.11141 [astro-ph.CO]}
  \BibitemShut {NoStop}%
\bibitem [{\citenamefont {Antonello}\ \emph {et~al.}(2019)\citenamefont
  {Antonello} \emph {et~al.}}]{Antonello:2018fvx}%
  \BibitemOpen
  \bibfield  {author} {\bibinfo {author} {\bibfnamefont {M.}~\bibnamefont
  {Antonello}} \emph {et~al.} (\bibinfo {collaboration} {SABRE}),\ }\href
  {https://doi.org/10.1140/epjc/s10052-019-6860-y} {\bibfield  {journal}
  {\bibinfo  {journal} {Eur. Phys. J. C}\ }\textbf {\bibinfo {volume} {79}},\
  \bibinfo {pages} {363} (\bibinfo {year} {2019})},\ \Eprint
  {https://arxiv.org/abs/1806.09340} {arXiv:1806.09340 [physics.ins-det]}
  \BibitemShut {NoStop}%
\bibitem [{\citenamefont {Agnese}\ \emph {et~al.}(2019)\citenamefont {Agnese}
  \emph {et~al.}}]{Agnese:2018gze}%
  \BibitemOpen
  \bibfield  {author} {\bibinfo {author} {\bibfnamefont {R.}~\bibnamefont
  {Agnese}} \emph {et~al.} (\bibinfo {collaboration} {SuperCDMS}),\ }\href
  {https://doi.org/10.1103/PhysRevD.99.062001} {\bibfield  {journal} {\bibinfo
  {journal} {Phys. Rev. D}\ }\textbf {\bibinfo {volume} {99}},\ \bibinfo
  {pages} {062001} (\bibinfo {year} {2019})},\ \Eprint
  {https://arxiv.org/abs/1808.09098} {arXiv:1808.09098 [astro-ph.CO]}
  \BibitemShut {NoStop}%
\bibitem [{\citenamefont {Alkhatib}\ \emph {et~al.}(2020)\citenamefont
  {Alkhatib} \emph {et~al.}}]{Alkhatib:2020slm}%
  \BibitemOpen
  \bibfield  {author} {\bibinfo {author} {\bibfnamefont {I.}~\bibnamefont
  {Alkhatib}} \emph {et~al.} (\bibinfo {collaboration} {SuperCDMS}),\
  }\href@noop {} {\  (\bibinfo {year} {2020})},\ \Eprint
  {https://arxiv.org/abs/2007.14289} {arXiv:2007.14289 [hep-ex]} \BibitemShut
  {NoStop}%
\bibitem [{\citenamefont {Aprile}\ \emph {et~al.}(2019)\citenamefont {Aprile}
  \emph {et~al.}}]{Aprile:2019xxb}%
  \BibitemOpen
  \bibfield  {author} {\bibinfo {author} {\bibfnamefont {E.}~\bibnamefont
  {Aprile}} \emph {et~al.} (\bibinfo {collaboration} {XENON}),\ }\href
  {https://doi.org/10.1103/PhysRevLett.123.251801} {\bibfield  {journal}
  {\bibinfo  {journal} {Phys. Rev. Lett.}\ }\textbf {\bibinfo {volume} {123}},\
  \bibinfo {pages} {251801} (\bibinfo {year} {2019})},\ \Eprint
  {https://arxiv.org/abs/1907.11485} {arXiv:1907.11485 [hep-ex]} \BibitemShut
  {NoStop}%
\bibitem [{\citenamefont {Hall}\ \emph {et~al.}(2010)\citenamefont {Hall},
  \citenamefont {Jedamzik}, \citenamefont {March-Russell},\ and\ \citenamefont
  {West}}]{Hall:2009bx}%
  \BibitemOpen
  \bibfield  {author} {\bibinfo {author} {\bibfnamefont {L.~J.}\ \bibnamefont
  {Hall}}, \bibinfo {author} {\bibfnamefont {K.}~\bibnamefont {Jedamzik}},
  \bibinfo {author} {\bibfnamefont {J.}~\bibnamefont {March-Russell}},\ and\
  \bibinfo {author} {\bibfnamefont {S.~M.}\ \bibnamefont {West}},\ }\href
  {https://doi.org/10.1007/JHEP03(2010)080} {\bibfield  {journal} {\bibinfo
  {journal} {JHEP}\ }\textbf {\bibinfo {volume} {03}},\ \bibinfo {pages}
  {080}},\ \Eprint {https://arxiv.org/abs/0911.1120} {arXiv:0911.1120 [hep-ph]}
  \BibitemShut {NoStop}%
\bibitem [{\citenamefont {Bernal}\ \emph {et~al.}(2017)\citenamefont {Bernal},
  \citenamefont {Heikinheimo}, \citenamefont {Tenkanen}, \citenamefont
  {Tuominen},\ and\ \citenamefont {Vaskonen}}]{Bernal:2017kxu}%
  \BibitemOpen
  \bibfield  {author} {\bibinfo {author} {\bibfnamefont {N.}~\bibnamefont
  {Bernal}}, \bibinfo {author} {\bibfnamefont {M.}~\bibnamefont {Heikinheimo}},
  \bibinfo {author} {\bibfnamefont {T.}~\bibnamefont {Tenkanen}}, \bibinfo
  {author} {\bibfnamefont {K.}~\bibnamefont {Tuominen}},\ and\ \bibinfo
  {author} {\bibfnamefont {V.}~\bibnamefont {Vaskonen}},\ }\href
  {https://doi.org/10.1142/S0217751X1730023X} {\bibfield  {journal} {\bibinfo
  {journal} {Int. J. Mod. Phys. A}\ }\textbf {\bibinfo {volume} {32}},\
  \bibinfo {pages} {1730023} (\bibinfo {year} {2017})},\ \Eprint
  {https://arxiv.org/abs/1706.07442} {arXiv:1706.07442 [hep-ph]} \BibitemShut
  {NoStop}%
\bibitem [{\citenamefont {Strassler}\ and\ \citenamefont
  {Zurek}(2007)}]{Strassler:2006im}%
  \BibitemOpen
  \bibfield  {author} {\bibinfo {author} {\bibfnamefont {M.~J.}\ \bibnamefont
  {Strassler}}\ and\ \bibinfo {author} {\bibfnamefont {K.~M.}\ \bibnamefont
  {Zurek}},\ }\href {https://doi.org/10.1016/j.physletb.2007.06.055} {\bibfield
   {journal} {\bibinfo  {journal} {Phys. Lett. B}\ }\textbf {\bibinfo {volume}
  {651}},\ \bibinfo {pages} {374} (\bibinfo {year} {2007})},\ \Eprint
  {https://arxiv.org/abs/hep-ph/0604261} {arXiv:hep-ph/0604261} \BibitemShut
  {NoStop}%
\bibitem [{\citenamefont {Arkani-Hamed}\ and\ \citenamefont
  {Weiner}(2008)}]{ArkaniHamed:2008qp}%
  \BibitemOpen
  \bibfield  {author} {\bibinfo {author} {\bibfnamefont {N.}~\bibnamefont
  {Arkani-Hamed}}\ and\ \bibinfo {author} {\bibfnamefont {N.}~\bibnamefont
  {Weiner}},\ }\href {https://doi.org/10.1088/1126-6708/2008/12/104} {\bibfield
   {journal} {\bibinfo  {journal} {JHEP}\ }\textbf {\bibinfo {volume} {12}},\
  \bibinfo {pages} {104}},\ \Eprint {https://arxiv.org/abs/0810.0714}
  {arXiv:0810.0714 [hep-ph]} \BibitemShut {NoStop}%
\bibitem [{\citenamefont {Cheung}\ \emph {et~al.}(2009)\citenamefont {Cheung},
  \citenamefont {Ruderman}, \citenamefont {Wang},\ and\ \citenamefont
  {Yavin}}]{Cheung:2009qd}%
  \BibitemOpen
  \bibfield  {author} {\bibinfo {author} {\bibfnamefont {C.}~\bibnamefont
  {Cheung}}, \bibinfo {author} {\bibfnamefont {J.~T.}\ \bibnamefont
  {Ruderman}}, \bibinfo {author} {\bibfnamefont {L.-T.}\ \bibnamefont {Wang}},\
  and\ \bibinfo {author} {\bibfnamefont {I.}~\bibnamefont {Yavin}},\ }\href
  {https://doi.org/10.1103/PhysRevD.80.035008} {\bibfield  {journal} {\bibinfo
  {journal} {Phys. Rev.}\ }\textbf {\bibinfo {volume} {D80}},\ \bibinfo {pages}
  {035008} (\bibinfo {year} {2009})},\ \Eprint
  {https://arxiv.org/abs/0902.3246} {arXiv:0902.3246 [hep-ph]} \BibitemShut
  {NoStop}%
\bibitem [{\citenamefont {Morrissey}\ \emph {et~al.}(2009)\citenamefont
  {Morrissey}, \citenamefont {Poland},\ and\ \citenamefont
  {Zurek}}]{Morrissey:2009ur}%
  \BibitemOpen
  \bibfield  {author} {\bibinfo {author} {\bibfnamefont {D.~E.}\ \bibnamefont
  {Morrissey}}, \bibinfo {author} {\bibfnamefont {D.}~\bibnamefont {Poland}},\
  and\ \bibinfo {author} {\bibfnamefont {K.~M.}\ \bibnamefont {Zurek}},\ }\href
  {https://doi.org/10.1088/1126-6708/2009/07/050} {\bibfield  {journal}
  {\bibinfo  {journal} {JHEP}\ }\textbf {\bibinfo {volume} {07}},\ \bibinfo
  {pages} {050}},\ \Eprint {https://arxiv.org/abs/0904.2567} {arXiv:0904.2567
  [hep-ph]} \BibitemShut {NoStop}%
\bibitem [{\citenamefont {Kaplan}\ \emph {et~al.}(2009)\citenamefont {Kaplan},
  \citenamefont {Luty},\ and\ \citenamefont {Zurek}}]{Kaplan:2009ag}%
  \BibitemOpen
  \bibfield  {author} {\bibinfo {author} {\bibfnamefont {D.~E.}\ \bibnamefont
  {Kaplan}}, \bibinfo {author} {\bibfnamefont {M.~A.}\ \bibnamefont {Luty}},\
  and\ \bibinfo {author} {\bibfnamefont {K.~M.}\ \bibnamefont {Zurek}},\ }\href
  {https://doi.org/10.1103/PhysRevD.79.115016} {\bibfield  {journal} {\bibinfo
  {journal} {Phys. Rev.}\ }\textbf {\bibinfo {volume} {D79}},\ \bibinfo {pages}
  {115016} (\bibinfo {year} {2009})},\ \Eprint
  {https://arxiv.org/abs/0901.4117} {arXiv:0901.4117 [hep-ph]} \BibitemShut
  {NoStop}%
\bibitem [{\citenamefont {Cohen}\ \emph {et~al.}(2010)\citenamefont {Cohen},
  \citenamefont {Phalen}, \citenamefont {Pierce},\ and\ \citenamefont
  {Zurek}}]{Cohen:2010kn}%
  \BibitemOpen
  \bibfield  {author} {\bibinfo {author} {\bibfnamefont {T.}~\bibnamefont
  {Cohen}}, \bibinfo {author} {\bibfnamefont {D.~J.}\ \bibnamefont {Phalen}},
  \bibinfo {author} {\bibfnamefont {A.}~\bibnamefont {Pierce}},\ and\ \bibinfo
  {author} {\bibfnamefont {K.~M.}\ \bibnamefont {Zurek}},\ }\href
  {https://doi.org/10.1103/PhysRevD.82.056001} {\bibfield  {journal} {\bibinfo
  {journal} {Phys. Rev.}\ }\textbf {\bibinfo {volume} {D82}},\ \bibinfo {pages}
  {056001} (\bibinfo {year} {2010})},\ \Eprint
  {https://arxiv.org/abs/1005.1655} {arXiv:1005.1655 [hep-ph]} \BibitemShut
  {NoStop}%
\bibitem [{\citenamefont {Petraki}\ and\ \citenamefont
  {Volkas}(2013)}]{Petraki:2013wwa}%
  \BibitemOpen
  \bibfield  {author} {\bibinfo {author} {\bibfnamefont {K.}~\bibnamefont
  {Petraki}}\ and\ \bibinfo {author} {\bibfnamefont {R.~R.}\ \bibnamefont
  {Volkas}},\ }\href {https://doi.org/10.1142/S0217751X13300287} {\bibfield
  {journal} {\bibinfo  {journal} {Int. J. Mod. Phys. A}\ }\textbf {\bibinfo
  {volume} {28}},\ \bibinfo {pages} {1330028} (\bibinfo {year} {2013})},\
  \Eprint {https://arxiv.org/abs/1305.4939} {arXiv:1305.4939 [hep-ph]}
  \BibitemShut {NoStop}%
\bibitem [{\citenamefont {Zurek}(2014)}]{Zurek:2013wia}%
  \BibitemOpen
  \bibfield  {author} {\bibinfo {author} {\bibfnamefont {K.~M.}\ \bibnamefont
  {Zurek}},\ }\href {https://doi.org/10.1016/j.physrep.2013.12.001} {\bibfield
  {journal} {\bibinfo  {journal} {Phys. Rept.}\ }\textbf {\bibinfo {volume}
  {537}},\ \bibinfo {pages} {91} (\bibinfo {year} {2014})},\ \Eprint
  {https://arxiv.org/abs/1308.0338} {arXiv:1308.0338 [hep-ph]} \BibitemShut
  {NoStop}%
\bibitem [{\citenamefont {Hochberg}\ \emph {et~al.}(2014)\citenamefont
  {Hochberg}, \citenamefont {Kuflik}, \citenamefont {Volansky},\ and\
  \citenamefont {Wacker}}]{Hochberg:2014dra}%
  \BibitemOpen
  \bibfield  {author} {\bibinfo {author} {\bibfnamefont {Y.}~\bibnamefont
  {Hochberg}}, \bibinfo {author} {\bibfnamefont {E.}~\bibnamefont {Kuflik}},
  \bibinfo {author} {\bibfnamefont {T.}~\bibnamefont {Volansky}},\ and\
  \bibinfo {author} {\bibfnamefont {J.~G.}\ \bibnamefont {Wacker}},\ }\href
  {https://doi.org/10.1103/PhysRevLett.113.171301} {\bibfield  {journal}
  {\bibinfo  {journal} {Phys. Rev. Lett.}\ }\textbf {\bibinfo {volume} {113}},\
  \bibinfo {pages} {171301} (\bibinfo {year} {2014})},\ \Eprint
  {https://arxiv.org/abs/1402.5143} {arXiv:1402.5143 [hep-ph]} \BibitemShut
  {NoStop}%
\bibitem [{\citenamefont {Hochberg}\ \emph {et~al.}(2015)\citenamefont
  {Hochberg}, \citenamefont {Kuflik}, \citenamefont {Murayama}, \citenamefont
  {Volansky},\ and\ \citenamefont {Wacker}}]{Hochberg:2014kqa}%
  \BibitemOpen
  \bibfield  {author} {\bibinfo {author} {\bibfnamefont {Y.}~\bibnamefont
  {Hochberg}}, \bibinfo {author} {\bibfnamefont {E.}~\bibnamefont {Kuflik}},
  \bibinfo {author} {\bibfnamefont {H.}~\bibnamefont {Murayama}}, \bibinfo
  {author} {\bibfnamefont {T.}~\bibnamefont {Volansky}},\ and\ \bibinfo
  {author} {\bibfnamefont {J.~G.}\ \bibnamefont {Wacker}},\ }\href
  {https://doi.org/10.1103/PhysRevLett.115.021301} {\bibfield  {journal}
  {\bibinfo  {journal} {Phys. Rev. Lett.}\ }\textbf {\bibinfo {volume} {115}},\
  \bibinfo {pages} {021301} (\bibinfo {year} {2015})},\ \Eprint
  {https://arxiv.org/abs/1411.3727} {arXiv:1411.3727 [hep-ph]} \BibitemShut
  {NoStop}%
\bibitem [{\citenamefont {Essig}\ \emph {et~al.}(2012)\citenamefont {Essig},
  \citenamefont {Mardon},\ and\ \citenamefont {Volansky}}]{Essig:2011nj}%
  \BibitemOpen
  \bibfield  {author} {\bibinfo {author} {\bibfnamefont {R.}~\bibnamefont
  {Essig}}, \bibinfo {author} {\bibfnamefont {J.}~\bibnamefont {Mardon}},\ and\
  \bibinfo {author} {\bibfnamefont {T.}~\bibnamefont {Volansky}},\ }\href
  {https://doi.org/10.1103/PhysRevD.85.076007} {\bibfield  {journal} {\bibinfo
  {journal} {Phys. Rev.}\ }\textbf {\bibinfo {volume} {D85}},\ \bibinfo {pages}
  {076007} (\bibinfo {year} {2012})},\ \Eprint
  {https://arxiv.org/abs/1108.5383} {arXiv:1108.5383 [hep-ph]} \BibitemShut
  {NoStop}%
\bibitem [{\citenamefont {Graham}\ \emph {et~al.}(2012)\citenamefont {Graham},
  \citenamefont {Kaplan}, \citenamefont {Rajendran},\ and\ \citenamefont
  {Walters}}]{Graham:2012su}%
  \BibitemOpen
  \bibfield  {author} {\bibinfo {author} {\bibfnamefont {P.~W.}\ \bibnamefont
  {Graham}}, \bibinfo {author} {\bibfnamefont {D.~E.}\ \bibnamefont {Kaplan}},
  \bibinfo {author} {\bibfnamefont {S.}~\bibnamefont {Rajendran}},\ and\
  \bibinfo {author} {\bibfnamefont {M.~T.}\ \bibnamefont {Walters}},\ }\href
  {https://doi.org/10.1016/j.dark.2012.09.001} {\bibfield  {journal} {\bibinfo
  {journal} {Phys. Dark Univ.}\ }\textbf {\bibinfo {volume} {1}},\ \bibinfo
  {pages} {32} (\bibinfo {year} {2012})},\ \Eprint
  {https://arxiv.org/abs/1203.2531} {arXiv:1203.2531 [hep-ph]} \BibitemShut
  {NoStop}%
\bibitem [{\citenamefont {Lee}\ \emph {et~al.}(2015)\citenamefont {Lee},
  \citenamefont {Lisanti}, \citenamefont {Mishra-Sharma},\ and\ \citenamefont
  {Safdi}}]{Lee:2015qva}%
  \BibitemOpen
  \bibfield  {author} {\bibinfo {author} {\bibfnamefont {S.~K.}\ \bibnamefont
  {Lee}}, \bibinfo {author} {\bibfnamefont {M.}~\bibnamefont {Lisanti}},
  \bibinfo {author} {\bibfnamefont {S.}~\bibnamefont {Mishra-Sharma}},\ and\
  \bibinfo {author} {\bibfnamefont {B.~R.}\ \bibnamefont {Safdi}},\ }\href
  {https://doi.org/10.1103/PhysRevD.92.083517} {\bibfield  {journal} {\bibinfo
  {journal} {Phys. Rev.}\ }\textbf {\bibinfo {volume} {D92}},\ \bibinfo {pages}
  {083517} (\bibinfo {year} {2015})},\ \Eprint
  {https://arxiv.org/abs/1508.07361} {arXiv:1508.07361 [hep-ph]} \BibitemShut
  {NoStop}%
\bibitem [{\citenamefont {Essig}\ \emph {et~al.}(2016)\citenamefont {Essig},
  \citenamefont {Fernandez-Serra}, \citenamefont {Mardon}, \citenamefont
  {Soto}, \citenamefont {Volansky},\ and\ \citenamefont {Yu}}]{Essig:2015cda}%
  \BibitemOpen
  \bibfield  {author} {\bibinfo {author} {\bibfnamefont {R.}~\bibnamefont
  {Essig}}, \bibinfo {author} {\bibfnamefont {M.}~\bibnamefont
  {Fernandez-Serra}}, \bibinfo {author} {\bibfnamefont {J.}~\bibnamefont
  {Mardon}}, \bibinfo {author} {\bibfnamefont {A.}~\bibnamefont {Soto}},
  \bibinfo {author} {\bibfnamefont {T.}~\bibnamefont {Volansky}},\ and\
  \bibinfo {author} {\bibfnamefont {T.-T.}\ \bibnamefont {Yu}},\ }\href
  {https://doi.org/10.1007/JHEP05(2016)046} {\bibfield  {journal} {\bibinfo
  {journal} {JHEP}\ }\textbf {\bibinfo {volume} {05}},\ \bibinfo {pages}
  {046}},\ \Eprint {https://arxiv.org/abs/1509.01598} {arXiv:1509.01598
  [hep-ph]} \BibitemShut {NoStop}%
\bibitem [{\citenamefont {Derenzo}\ \emph {et~al.}(2017)\citenamefont
  {Derenzo}, \citenamefont {Essig}, \citenamefont {Massari}, \citenamefont
  {Soto},\ and\ \citenamefont {Yu}}]{Derenzo:2016fse}%
  \BibitemOpen
  \bibfield  {author} {\bibinfo {author} {\bibfnamefont {S.}~\bibnamefont
  {Derenzo}}, \bibinfo {author} {\bibfnamefont {R.}~\bibnamefont {Essig}},
  \bibinfo {author} {\bibfnamefont {A.}~\bibnamefont {Massari}}, \bibinfo
  {author} {\bibfnamefont {A.}~\bibnamefont {Soto}},\ and\ \bibinfo {author}
  {\bibfnamefont {T.-T.}\ \bibnamefont {Yu}},\ }\href
  {https://doi.org/10.1103/PhysRevD.96.016026} {\bibfield  {journal} {\bibinfo
  {journal} {Phys. Rev.}\ }\textbf {\bibinfo {volume} {D96}},\ \bibinfo {pages}
  {016026} (\bibinfo {year} {2017})},\ \Eprint
  {https://arxiv.org/abs/1607.01009} {arXiv:1607.01009 [hep-ph]} \BibitemShut
  {NoStop}%
\bibitem [{\citenamefont {Hochberg}\ \emph {et~al.}(2017)\citenamefont
  {Hochberg}, \citenamefont {Kahn}, \citenamefont {Lisanti}, \citenamefont
  {Tully},\ and\ \citenamefont {Zurek}}]{Hochberg:2016ntt}%
  \BibitemOpen
  \bibfield  {author} {\bibinfo {author} {\bibfnamefont {Y.}~\bibnamefont
  {Hochberg}}, \bibinfo {author} {\bibfnamefont {Y.}~\bibnamefont {Kahn}},
  \bibinfo {author} {\bibfnamefont {M.}~\bibnamefont {Lisanti}}, \bibinfo
  {author} {\bibfnamefont {C.~G.}\ \bibnamefont {Tully}},\ and\ \bibinfo
  {author} {\bibfnamefont {K.~M.}\ \bibnamefont {Zurek}},\ }\href
  {https://doi.org/10.1016/j.physletb.2017.06.051} {\bibfield  {journal}
  {\bibinfo  {journal} {Phys. Lett.}\ }\textbf {\bibinfo {volume} {B772}},\
  \bibinfo {pages} {239} (\bibinfo {year} {2017})},\ \Eprint
  {https://arxiv.org/abs/1606.08849} {arXiv:1606.08849 [hep-ph]} \BibitemShut
  {NoStop}%
\bibitem [{\citenamefont {Essig}\ \emph {et~al.}(2017)\citenamefont {Essig},
  \citenamefont {Volansky},\ and\ \citenamefont {Yu}}]{Essig:2017kqs}%
  \BibitemOpen
  \bibfield  {author} {\bibinfo {author} {\bibfnamefont {R.}~\bibnamefont
  {Essig}}, \bibinfo {author} {\bibfnamefont {T.}~\bibnamefont {Volansky}},\
  and\ \bibinfo {author} {\bibfnamefont {T.-T.}\ \bibnamefont {Yu}},\ }\href
  {https://doi.org/10.1103/PhysRevD.96.043017} {\bibfield  {journal} {\bibinfo
  {journal} {Phys. Rev.}\ }\textbf {\bibinfo {volume} {D96}},\ \bibinfo {pages}
  {043017} (\bibinfo {year} {2017})},\ \Eprint
  {https://arxiv.org/abs/1703.00910} {arXiv:1703.00910 [hep-ph]} \BibitemShut
  {NoStop}%
\bibitem [{\citenamefont {Kurinsky}\ \emph {et~al.}(2019)\citenamefont
  {Kurinsky}, \citenamefont {Yu}, \citenamefont {Hochberg},\ and\ \citenamefont
  {Cabrera}}]{Kurinsky:2019pgb}%
  \BibitemOpen
  \bibfield  {author} {\bibinfo {author} {\bibfnamefont {N.~A.}\ \bibnamefont
  {Kurinsky}}, \bibinfo {author} {\bibfnamefont {T.~C.}\ \bibnamefont {Yu}},
  \bibinfo {author} {\bibfnamefont {Y.}~\bibnamefont {Hochberg}},\ and\
  \bibinfo {author} {\bibfnamefont {B.}~\bibnamefont {Cabrera}},\ }\href
  {https://doi.org/10.1103/PhysRevD.99.123005} {\bibfield  {journal} {\bibinfo
  {journal} {Phys. Rev.}\ }\textbf {\bibinfo {volume} {D99}},\ \bibinfo {pages}
  {123005} (\bibinfo {year} {2019})},\ \Eprint
  {https://arxiv.org/abs/1901.07569} {arXiv:1901.07569 [hep-ex]} \BibitemShut
  {NoStop}%
\bibitem [{\citenamefont {Blanco}\ \emph {et~al.}(2020)\citenamefont {Blanco},
  \citenamefont {Collar}, \citenamefont {Kahn},\ and\ \citenamefont
  {Lillard}}]{Blanco:2019lrf}%
  \BibitemOpen
  \bibfield  {author} {\bibinfo {author} {\bibfnamefont {C.}~\bibnamefont
  {Blanco}}, \bibinfo {author} {\bibfnamefont {J.~I.}\ \bibnamefont {Collar}},
  \bibinfo {author} {\bibfnamefont {Y.}~\bibnamefont {Kahn}},\ and\ \bibinfo
  {author} {\bibfnamefont {B.}~\bibnamefont {Lillard}},\ }\href
  {https://doi.org/10.1103/PhysRevD.101.056001} {\bibfield  {journal} {\bibinfo
   {journal} {Phys. Rev. D}\ }\textbf {\bibinfo {volume} {101}},\ \bibinfo
  {pages} {056001} (\bibinfo {year} {2020})},\ \Eprint
  {https://arxiv.org/abs/1912.02822} {arXiv:1912.02822 [hep-ph]} \BibitemShut
  {NoStop}%
\bibitem [{\citenamefont {Catena}\ \emph {et~al.}(2020)\citenamefont {Catena},
  \citenamefont {Emken}, \citenamefont {Spaldin},\ and\ \citenamefont
  {Tarantino}}]{Catena:2019gfa}%
  \BibitemOpen
  \bibfield  {author} {\bibinfo {author} {\bibfnamefont {R.}~\bibnamefont
  {Catena}}, \bibinfo {author} {\bibfnamefont {T.}~\bibnamefont {Emken}},
  \bibinfo {author} {\bibfnamefont {N.~A.}\ \bibnamefont {Spaldin}},\ and\
  \bibinfo {author} {\bibfnamefont {W.}~\bibnamefont {Tarantino}},\ }\href
  {https://doi.org/10.1103/PhysRevResearch.2.033195} {\bibfield  {journal}
  {\bibinfo  {journal} {Phys. Rev. Res.}\ }\textbf {\bibinfo {volume} {2}},\
  \bibinfo {pages} {033195} (\bibinfo {year} {2020})},\ \Eprint
  {https://arxiv.org/abs/1912.08204} {arXiv:1912.08204 [hep-ph]} \BibitemShut
  {NoStop}%
\bibitem [{\citenamefont {Settimo}(2020)}]{Settimo:2020cbq}%
  \BibitemOpen
  \bibfield  {author} {\bibinfo {author} {\bibfnamefont {M.}~\bibnamefont
  {Settimo}} (\bibinfo {collaboration} {DAMIC, DAMIC-M}),\ }in\ \href@noop {}
  {\emph {\bibinfo {booktitle} {{16th Rencontres du Vietnam}: {Theory meeting
  experiment: Particle Astrophysics and Cosmology}}}}\ (\bibinfo {year}
  {2020})\ \Eprint {https://arxiv.org/abs/2003.09497} {arXiv:2003.09497
  [hep-ex]} \BibitemShut {NoStop}%
\bibitem [{\citenamefont {Barak}\ \emph {et~al.}(2020)\citenamefont {Barak}
  \emph {et~al.}}]{Barak:2020fql}%
  \BibitemOpen
  \bibfield  {author} {\bibinfo {author} {\bibfnamefont {L.}~\bibnamefont
  {Barak}} \emph {et~al.} (\bibinfo {collaboration} {SENSEI}),\ }\href
  {https://doi.org/10.1103/PhysRevLett.125.171802} {\bibfield  {journal}
  {\bibinfo  {journal} {Phys. Rev. Lett.}\ }\textbf {\bibinfo {volume} {125}},\
  \bibinfo {pages} {171802} (\bibinfo {year} {2020})},\ \Eprint
  {https://arxiv.org/abs/2004.11378} {arXiv:2004.11378 [astro-ph.CO]}
  \BibitemShut {NoStop}%
\bibitem [{\citenamefont {Amaral}\ \emph {et~al.}(2020)\citenamefont {Amaral}
  \emph {et~al.}}]{Amaral:2020ryn}%
  \BibitemOpen
  \bibfield  {author} {\bibinfo {author} {\bibfnamefont {D.~W.}\ \bibnamefont
  {Amaral}} \emph {et~al.} (\bibinfo {collaboration} {SuperCDMS}),\ }\href
  {https://doi.org/10.1103/PhysRevD.102.091101} {\bibfield  {journal} {\bibinfo
   {journal} {Phys. Rev. D}\ }\textbf {\bibinfo {volume} {102}},\ \bibinfo
  {pages} {091101} (\bibinfo {year} {2020})},\ \Eprint
  {https://arxiv.org/abs/2005.14067} {arXiv:2005.14067 [hep-ex]} \BibitemShut
  {NoStop}%
\bibitem [{\citenamefont {Schutz}\ and\ \citenamefont
  {Zurek}(2016)}]{Schutz:2016tid}%
  \BibitemOpen
  \bibfield  {author} {\bibinfo {author} {\bibfnamefont {K.}~\bibnamefont
  {Schutz}}\ and\ \bibinfo {author} {\bibfnamefont {K.~M.}\ \bibnamefont
  {Zurek}},\ }\href {https://doi.org/10.1103/PhysRevLett.117.121302} {\bibfield
   {journal} {\bibinfo  {journal} {Phys. Rev. Lett.}\ }\textbf {\bibinfo
  {volume} {117}},\ \bibinfo {pages} {121302} (\bibinfo {year} {2016})},\
  \Eprint {https://arxiv.org/abs/1604.08206} {arXiv:1604.08206 [hep-ph]}
  \BibitemShut {NoStop}%
\bibitem [{\citenamefont {Knapen}\ \emph {et~al.}(2017)\citenamefont {Knapen},
  \citenamefont {Lin},\ and\ \citenamefont {Zurek}}]{Knapen:2016cue}%
  \BibitemOpen
  \bibfield  {author} {\bibinfo {author} {\bibfnamefont {S.}~\bibnamefont
  {Knapen}}, \bibinfo {author} {\bibfnamefont {T.}~\bibnamefont {Lin}},\ and\
  \bibinfo {author} {\bibfnamefont {K.~M.}\ \bibnamefont {Zurek}},\ }\href
  {https://doi.org/10.1103/PhysRevD.95.056019} {\bibfield  {journal} {\bibinfo
  {journal} {Phys. Rev.}\ }\textbf {\bibinfo {volume} {D95}},\ \bibinfo {pages}
  {056019} (\bibinfo {year} {2017})},\ \Eprint
  {https://arxiv.org/abs/1611.06228} {arXiv:1611.06228 [hep-ph]} \BibitemShut
  {NoStop}%
\bibitem [{\citenamefont {Acanfora}\ \emph {et~al.}(2019)\citenamefont
  {Acanfora}, \citenamefont {Esposito},\ and\ \citenamefont
  {Polosa}}]{Acanfora:2019con}%
  \BibitemOpen
  \bibfield  {author} {\bibinfo {author} {\bibfnamefont {F.}~\bibnamefont
  {Acanfora}}, \bibinfo {author} {\bibfnamefont {A.}~\bibnamefont {Esposito}},\
  and\ \bibinfo {author} {\bibfnamefont {A.~D.}\ \bibnamefont {Polosa}},\
  }\href {https://doi.org/10.1140/epjc/s10052-019-7057-0} {\bibfield  {journal}
  {\bibinfo  {journal} {Eur. Phys. J.}\ }\textbf {\bibinfo {volume} {C79}},\
  \bibinfo {pages} {549} (\bibinfo {year} {2019})},\ \Eprint
  {https://arxiv.org/abs/1902.02361} {arXiv:1902.02361 [hep-ph]} \BibitemShut
  {NoStop}%
\bibitem [{\citenamefont {Caputo}\ \emph {et~al.}(2019)\citenamefont {Caputo},
  \citenamefont {Esposito},\ and\ \citenamefont {Polosa}}]{Caputo:2019cyg}%
  \BibitemOpen
  \bibfield  {author} {\bibinfo {author} {\bibfnamefont {A.}~\bibnamefont
  {Caputo}}, \bibinfo {author} {\bibfnamefont {A.}~\bibnamefont {Esposito}},\
  and\ \bibinfo {author} {\bibfnamefont {A.~D.}\ \bibnamefont {Polosa}},\
  }\href {https://doi.org/10.1103/PhysRevD.100.116007} {\bibfield  {journal}
  {\bibinfo  {journal} {Phys. Rev. D}\ }\textbf {\bibinfo {volume} {100}},\
  \bibinfo {pages} {116007} (\bibinfo {year} {2019})},\ \Eprint
  {https://arxiv.org/abs/1907.10635} {arXiv:1907.10635 [hep-ph]} \BibitemShut
  {NoStop}%
\bibitem [{\citenamefont {Baym}\ \emph {et~al.}(2020)\citenamefont {Baym},
  \citenamefont {Beck}, \citenamefont {Filippini}, \citenamefont {Pethick},\
  and\ \citenamefont {Shelton}}]{Baym:2020uos}%
  \BibitemOpen
  \bibfield  {author} {\bibinfo {author} {\bibfnamefont {G.}~\bibnamefont
  {Baym}}, \bibinfo {author} {\bibfnamefont {D.}~\bibnamefont {Beck}}, \bibinfo
  {author} {\bibfnamefont {J.~P.}\ \bibnamefont {Filippini}}, \bibinfo {author}
  {\bibfnamefont {C.}~\bibnamefont {Pethick}},\ and\ \bibinfo {author}
  {\bibfnamefont {J.}~\bibnamefont {Shelton}},\ }\href
  {https://doi.org/10.1103/PhysRevD.102.035014} {\bibfield  {journal} {\bibinfo
   {journal} {Phys. Rev. D}\ }\textbf {\bibinfo {volume} {102}},\ \bibinfo
  {pages} {035014} (\bibinfo {year} {2020})},\ \Eprint
  {https://arxiv.org/abs/2005.08824} {arXiv:2005.08824 [hep-ph]} \BibitemShut
  {NoStop}%
\bibitem [{\citenamefont {Caputo}\ \emph {et~al.}(2020)\citenamefont {Caputo},
  \citenamefont {Esposito}, \citenamefont {Piccinini}, \citenamefont {Polosa},\
  and\ \citenamefont {Rossi}}]{Caputo:2020sys}%
  \BibitemOpen
  \bibfield  {author} {\bibinfo {author} {\bibfnamefont {A.}~\bibnamefont
  {Caputo}}, \bibinfo {author} {\bibfnamefont {A.}~\bibnamefont {Esposito}},
  \bibinfo {author} {\bibfnamefont {F.}~\bibnamefont {Piccinini}}, \bibinfo
  {author} {\bibfnamefont {A.~D.}\ \bibnamefont {Polosa}},\ and\ \bibinfo
  {author} {\bibfnamefont {G.}~\bibnamefont {Rossi}},\ }\href@noop {} {\
  (\bibinfo {year} {2020})},\ \Eprint {https://arxiv.org/abs/2012.01432}
  {arXiv:2012.01432 [hep-ph]} \BibitemShut {NoStop}%
\bibitem [{\citenamefont {Knapen}\ \emph {et~al.}(2018)\citenamefont {Knapen},
  \citenamefont {Lin}, \citenamefont {Pyle},\ and\ \citenamefont
  {Zurek}}]{Knapen:2017ekk}%
  \BibitemOpen
  \bibfield  {author} {\bibinfo {author} {\bibfnamefont {S.}~\bibnamefont
  {Knapen}}, \bibinfo {author} {\bibfnamefont {T.}~\bibnamefont {Lin}},
  \bibinfo {author} {\bibfnamefont {M.}~\bibnamefont {Pyle}},\ and\ \bibinfo
  {author} {\bibfnamefont {K.~M.}\ \bibnamefont {Zurek}},\ }\href
  {https://doi.org/10.1016/j.physletb.2018.08.064} {\bibfield  {journal}
  {\bibinfo  {journal} {Phys. Lett.}\ }\textbf {\bibinfo {volume} {B785}},\
  \bibinfo {pages} {386} (\bibinfo {year} {2018})},\ \Eprint
  {https://arxiv.org/abs/1712.06598} {arXiv:1712.06598 [hep-ph]} \BibitemShut
  {NoStop}%
\bibitem [{\citenamefont {Griffin}\ \emph {et~al.}(2018)\citenamefont
  {Griffin}, \citenamefont {Knapen}, \citenamefont {Lin},\ and\ \citenamefont
  {Zurek}}]{Griffin:2018bjn}%
  \BibitemOpen
  \bibfield  {author} {\bibinfo {author} {\bibfnamefont {S.}~\bibnamefont
  {Griffin}}, \bibinfo {author} {\bibfnamefont {S.}~\bibnamefont {Knapen}},
  \bibinfo {author} {\bibfnamefont {T.}~\bibnamefont {Lin}},\ and\ \bibinfo
  {author} {\bibfnamefont {K.~M.}\ \bibnamefont {Zurek}},\ }\href
  {https://doi.org/10.1103/PhysRevD.98.115034} {\bibfield  {journal} {\bibinfo
  {journal} {Phys. Rev.}\ }\textbf {\bibinfo {volume} {D98}},\ \bibinfo {pages}
  {115034} (\bibinfo {year} {2018})},\ \Eprint
  {https://arxiv.org/abs/1807.10291} {arXiv:1807.10291 [hep-ph]} \BibitemShut
  {NoStop}%
\bibitem [{\citenamefont {Trickle}\ \emph
  {et~al.}(2020{\natexlab{a}})\citenamefont {Trickle}, \citenamefont {Zhang},
  \citenamefont {Zurek}, \citenamefont {Inzani},\ and\ \citenamefont
  {Griffin}}]{Trickle:2019nya}%
  \BibitemOpen
  \bibfield  {author} {\bibinfo {author} {\bibfnamefont {T.}~\bibnamefont
  {Trickle}}, \bibinfo {author} {\bibfnamefont {Z.}~\bibnamefont {Zhang}},
  \bibinfo {author} {\bibfnamefont {K.~M.}\ \bibnamefont {Zurek}}, \bibinfo
  {author} {\bibfnamefont {K.}~\bibnamefont {Inzani}},\ and\ \bibinfo {author}
  {\bibfnamefont {S.}~\bibnamefont {Griffin}},\ }\href
  {https://doi.org/10.1007/JHEP03(2020)036} {\bibfield  {journal} {\bibinfo
  {journal} {JHEP}\ }\textbf {\bibinfo {volume} {03}},\ \bibinfo {pages}
  {036}},\ \Eprint {https://arxiv.org/abs/1910.08092} {arXiv:1910.08092
  [hep-ph]} \BibitemShut {NoStop}%
\bibitem [{\citenamefont {Griffin}\ \emph
  {et~al.}(2020{\natexlab{a}})\citenamefont {Griffin}, \citenamefont {Inzani},
  \citenamefont {Trickle}, \citenamefont {Zhang},\ and\ \citenamefont
  {Zurek}}]{Griffin:2019mvc}%
  \BibitemOpen
  \bibfield  {author} {\bibinfo {author} {\bibfnamefont {S.~M.}\ \bibnamefont
  {Griffin}}, \bibinfo {author} {\bibfnamefont {K.}~\bibnamefont {Inzani}},
  \bibinfo {author} {\bibfnamefont {T.}~\bibnamefont {Trickle}}, \bibinfo
  {author} {\bibfnamefont {Z.}~\bibnamefont {Zhang}},\ and\ \bibinfo {author}
  {\bibfnamefont {K.~M.}\ \bibnamefont {Zurek}},\ }\href
  {https://doi.org/10.1103/PhysRevD.101.055004} {\bibfield  {journal} {\bibinfo
   {journal} {Phys. Rev. D}\ }\textbf {\bibinfo {volume} {101}},\ \bibinfo
  {pages} {055004} (\bibinfo {year} {2020}{\natexlab{a}})},\ \Eprint
  {https://arxiv.org/abs/1910.10716} {arXiv:1910.10716 [hep-ph]} \BibitemShut
  {NoStop}%
\bibitem [{\citenamefont {Campbell-Deem}\ \emph {et~al.}(2020)\citenamefont
  {Campbell-Deem}, \citenamefont {Cox}, \citenamefont {Knapen}, \citenamefont
  {Lin},\ and\ \citenamefont {Melia}}]{Campbell-Deem:2019hdx}%
  \BibitemOpen
  \bibfield  {author} {\bibinfo {author} {\bibfnamefont {B.}~\bibnamefont
  {Campbell-Deem}}, \bibinfo {author} {\bibfnamefont {P.}~\bibnamefont {Cox}},
  \bibinfo {author} {\bibfnamefont {S.}~\bibnamefont {Knapen}}, \bibinfo
  {author} {\bibfnamefont {T.}~\bibnamefont {Lin}},\ and\ \bibinfo {author}
  {\bibfnamefont {T.}~\bibnamefont {Melia}},\ }\href
  {https://doi.org/10.1103/PhysRevD.101.036006} {\bibfield  {journal} {\bibinfo
   {journal} {Phys. Rev. D}\ }\textbf {\bibinfo {volume} {101}},\ \bibinfo
  {pages} {036006} (\bibinfo {year} {2020})},\ \bibinfo {note} {[Erratum:
  Phys.Rev.D 102, 019904 (2020)]},\ \Eprint {https://arxiv.org/abs/1911.03482}
  {arXiv:1911.03482 [hep-ph]} \BibitemShut {NoStop}%
\bibitem [{\citenamefont {Griffin}\ \emph
  {et~al.}(2020{\natexlab{b}})\citenamefont {Griffin}, \citenamefont
  {Hochberg}, \citenamefont {Inzani}, \citenamefont {Kurinsky}, \citenamefont
  {Lin},\ and\ \citenamefont {Yu}}]{Griffin:2020lgd}%
  \BibitemOpen
  \bibfield  {author} {\bibinfo {author} {\bibfnamefont {S.~M.}\ \bibnamefont
  {Griffin}}, \bibinfo {author} {\bibfnamefont {Y.}~\bibnamefont {Hochberg}},
  \bibinfo {author} {\bibfnamefont {K.}~\bibnamefont {Inzani}}, \bibinfo
  {author} {\bibfnamefont {N.}~\bibnamefont {Kurinsky}}, \bibinfo {author}
  {\bibfnamefont {T.}~\bibnamefont {Lin}},\ and\ \bibinfo {author}
  {\bibfnamefont {T.~C.}\ \bibnamefont {Yu}},\ }\href@noop {} {\  (\bibinfo
  {year} {2020}{\natexlab{b}})},\ \Eprint {https://arxiv.org/abs/2008.08560}
  {arXiv:2008.08560 [hep-ph]} \BibitemShut {NoStop}%
\bibitem [{\citenamefont {Trickle}\ \emph
  {et~al.}(2020{\natexlab{b}})\citenamefont {Trickle}, \citenamefont {Zhang},\
  and\ \citenamefont {Zurek}}]{Trickle:2020oki}%
  \BibitemOpen
  \bibfield  {author} {\bibinfo {author} {\bibfnamefont {T.}~\bibnamefont
  {Trickle}}, \bibinfo {author} {\bibfnamefont {Z.}~\bibnamefont {Zhang}},\
  and\ \bibinfo {author} {\bibfnamefont {K.~M.}\ \bibnamefont {Zurek}},\
  }\href@noop {} {\  (\bibinfo {year} {2020}{\natexlab{b}})},\ \Eprint
  {https://arxiv.org/abs/2009.13534} {arXiv:2009.13534 [hep-ph]} \BibitemShut
  {NoStop}%
\bibitem [{\citenamefont {Kahn}\ \emph {et~al.}(2020)\citenamefont {Kahn},
  \citenamefont {Krnjaic},\ and\ \citenamefont {Mandava}}]{Kahn:2020fef}%
  \BibitemOpen
  \bibfield  {author} {\bibinfo {author} {\bibfnamefont {Y.}~\bibnamefont
  {Kahn}}, \bibinfo {author} {\bibfnamefont {G.}~\bibnamefont {Krnjaic}},\ and\
  \bibinfo {author} {\bibfnamefont {B.}~\bibnamefont {Mandava}},\ }\href@noop
  {} {\  (\bibinfo {year} {2020})},\ \Eprint {https://arxiv.org/abs/2011.09477}
  {arXiv:2011.09477 [hep-ph]} \BibitemShut {NoStop}%
\bibitem [{\citenamefont {Knapen}\ \emph {et~al.}(2020)\citenamefont {Knapen},
  \citenamefont {Kozaczuk},\ and\ \citenamefont {Lin}}]{Knapen:2020aky}%
  \BibitemOpen
  \bibfield  {author} {\bibinfo {author} {\bibfnamefont {S.}~\bibnamefont
  {Knapen}}, \bibinfo {author} {\bibfnamefont {J.}~\bibnamefont {Kozaczuk}},\
  and\ \bibinfo {author} {\bibfnamefont {T.}~\bibnamefont {Lin}},\ }\href@noop
  {} {\  (\bibinfo {year} {2020})},\ \Eprint {https://arxiv.org/abs/2011.09496}
  {arXiv:2011.09496 [hep-ph]} \BibitemShut {NoStop}%
\bibitem [{\citenamefont {Pyle}\ \emph {et~al.}(2015)\citenamefont {Pyle},
  \citenamefont {Feliciano-Figueroa},\ and\ \citenamefont
  {Sadoulet}}]{Pyle:2015pya}%
  \BibitemOpen
  \bibfield  {author} {\bibinfo {author} {\bibfnamefont {M.}~\bibnamefont
  {Pyle}}, \bibinfo {author} {\bibfnamefont {E.}~\bibnamefont
  {Feliciano-Figueroa}},\ and\ \bibinfo {author} {\bibfnamefont
  {B.}~\bibnamefont {Sadoulet}},\ }\href@noop {} {\  (\bibinfo {year}
  {2015})},\ \Eprint {https://arxiv.org/abs/1503.01200} {arXiv:1503.01200
  [astro-ph.IM]} \BibitemShut {NoStop}%
\bibitem [{\citenamefont {Maris}\ \emph {et~al.}(2017)\citenamefont {Maris},
  \citenamefont {Seidel},\ and\ \citenamefont {Stein}}]{Maris:2017xvi}%
  \BibitemOpen
  \bibfield  {author} {\bibinfo {author} {\bibfnamefont {H.~J.}\ \bibnamefont
  {Maris}}, \bibinfo {author} {\bibfnamefont {G.~M.}\ \bibnamefont {Seidel}},\
  and\ \bibinfo {author} {\bibfnamefont {D.}~\bibnamefont {Stein}},\ }\href
  {https://doi.org/10.1103/PhysRevLett.119.181303} {\bibfield  {journal}
  {\bibinfo  {journal} {Phys. Rev. Lett.}\ }\textbf {\bibinfo {volume} {119}},\
  \bibinfo {pages} {181303} (\bibinfo {year} {2017})},\ \Eprint
  {https://arxiv.org/abs/1706.00117} {arXiv:1706.00117 [astro-ph.IM]}
  \BibitemShut {NoStop}%
\bibitem [{\citenamefont {Rothe}\ \emph {et~al.}(2018)\citenamefont {Rothe}
  \emph {et~al.}}]{Rothe:2018bnc}%
  \BibitemOpen
  \bibfield  {author} {\bibinfo {author} {\bibfnamefont {J.}~\bibnamefont
  {Rothe}} \emph {et~al.},\ }\href {https://doi.org/10.1007/s10909-018-1944-x}
  {\bibfield  {journal} {\bibinfo  {journal} {J. Low Temp. Phys.}\ }\textbf
  {\bibinfo {volume} {193}},\ \bibinfo {pages} {1160} (\bibinfo {year}
  {2018})}\BibitemShut {NoStop}%
\bibitem [{\citenamefont {Colantoni}\ \emph {et~al.}(2020)\citenamefont
  {Colantoni} \emph {et~al.}}]{Colantoni:2020cet}%
  \BibitemOpen
  \bibfield  {author} {\bibinfo {author} {\bibfnamefont {I.}~\bibnamefont
  {Colantoni}} \emph {et~al.},\ }\href
  {https://doi.org/10.1007/s10909-020-02408-3} {\bibfield  {journal} {\bibinfo
  {journal} {J. Low Temp. Phys.}\ }\textbf {\bibinfo {volume} {199}},\ \bibinfo
  {pages} {593} (\bibinfo {year} {2020})}\BibitemShut {NoStop}%
\bibitem [{\citenamefont {Fink}\ \emph {et~al.}(2020)\citenamefont {Fink} \emph
  {et~al.}}]{Fink:2020noh}%
  \BibitemOpen
  \bibfield  {author} {\bibinfo {author} {\bibfnamefont {C.}~\bibnamefont
  {Fink}} \emph {et~al.},\ }\href {https://doi.org/10.1063/5.0011130}
  {\bibfield  {journal} {\bibinfo  {journal} {AIP Adv.}\ }\textbf {\bibinfo
  {volume} {10}},\ \bibinfo {pages} {085221} (\bibinfo {year} {2020})},\
  \Eprint {https://arxiv.org/abs/2004.10257} {arXiv:2004.10257
  [physics.ins-det]} \BibitemShut {NoStop}%
\bibitem [{\citenamefont {Chang}\ \emph {et~al.}(2020)\citenamefont {Chang}
  \emph {et~al.}}]{tesseract}%
  \BibitemOpen
  \bibfield  {author} {\bibinfo {author} {\bibfnamefont {C.}~\bibnamefont
  {Chang}} \emph {et~al.}} (\bibinfo {year} {2020})\BibitemShut {NoStop}%
\bibitem [{\citenamefont {Du}\ \emph {et~al.}(2020)\citenamefont {Du},
  \citenamefont {Egana-Ugrinovic}, \citenamefont {Essig},\ and\ \citenamefont
  {Sholapurkar}}]{Du:2020ldo}%
  \BibitemOpen
  \bibfield  {author} {\bibinfo {author} {\bibfnamefont {P.}~\bibnamefont
  {Du}}, \bibinfo {author} {\bibfnamefont {D.}~\bibnamefont {Egana-Ugrinovic}},
  \bibinfo {author} {\bibfnamefont {R.}~\bibnamefont {Essig}},\ and\ \bibinfo
  {author} {\bibfnamefont {M.}~\bibnamefont {Sholapurkar}},\ }\href@noop {} {\
  (\bibinfo {year} {2020})},\ \Eprint {https://arxiv.org/abs/2011.13939}
  {arXiv:2011.13939 [hep-ph]} \BibitemShut {NoStop}%
\bibitem [{\citenamefont {Drukier}\ \emph {et~al.}(1986)\citenamefont
  {Drukier}, \citenamefont {Freese},\ and\ \citenamefont
  {Spergel}}]{Drukier:1986tm}%
  \BibitemOpen
  \bibfield  {author} {\bibinfo {author} {\bibfnamefont {A.~K.}\ \bibnamefont
  {Drukier}}, \bibinfo {author} {\bibfnamefont {K.}~\bibnamefont {Freese}},\
  and\ \bibinfo {author} {\bibfnamefont {D.~N.}\ \bibnamefont {Spergel}},\
  }\href {https://doi.org/10.1103/PhysRevD.33.3495} {\bibfield  {journal}
  {\bibinfo  {journal} {Phys. Rev. D}\ }\textbf {\bibinfo {volume} {33}},\
  \bibinfo {pages} {3495} (\bibinfo {year} {1986})}\BibitemShut {NoStop}%
\bibitem [{\citenamefont {Coskuner}\ \emph {et~al.}(2019)\citenamefont
  {Coskuner}, \citenamefont {Mitridate}, \citenamefont {Olivares},\ and\
  \citenamefont {Zurek}}]{Coskuner:2019odd}%
  \BibitemOpen
  \bibfield  {author} {\bibinfo {author} {\bibfnamefont {A.}~\bibnamefont
  {Coskuner}}, \bibinfo {author} {\bibfnamefont {A.}~\bibnamefont {Mitridate}},
  \bibinfo {author} {\bibfnamefont {A.}~\bibnamefont {Olivares}},\ and\
  \bibinfo {author} {\bibfnamefont {K.~M.}\ \bibnamefont {Zurek}},\ }\href@noop
  {} {\  (\bibinfo {year} {2019})},\ \Eprint {https://arxiv.org/abs/1909.09170}
  {arXiv:1909.09170 [hep-ph]} \BibitemShut {NoStop}%
\bibitem [{\citenamefont {Geilhufe}\ \emph {et~al.}(2019)\citenamefont
  {Geilhufe}, \citenamefont {Kahlhoefer},\ and\ \citenamefont
  {Winkler}}]{Geilhufe:2019ndy}%
  \BibitemOpen
  \bibfield  {author} {\bibinfo {author} {\bibfnamefont {R.~M.}\ \bibnamefont
  {Geilhufe}}, \bibinfo {author} {\bibfnamefont {F.}~\bibnamefont
  {Kahlhoefer}},\ and\ \bibinfo {author} {\bibfnamefont {M.~W.}\ \bibnamefont
  {Winkler}},\ }\href@noop {} {\  (\bibinfo {year} {2019})},\ \Eprint
  {https://arxiv.org/abs/1910.02091} {arXiv:1910.02091 [hep-ph]} \BibitemShut
  {NoStop}%
\bibitem [{dm-()}]{dm-phonon-scatter}%
  \BibitemOpen
  \href@noop {} {}\bibinfo {howpublished}
  {\url{\codelink}~\href{\codelink}{\faGithub}}\BibitemShut {NoStop}%
\bibitem [{dem()}]{demo}%
  \BibitemOpen
  \href@noop {} {}\bibinfo {howpublished} {\url{\demolink}.}\BibitemShut
  {Stop}%
\bibitem [{\citenamefont {Togo}()}]{phonondb}%
  \BibitemOpen
  \bibfield  {author} {\bibinfo {author} {\bibfnamefont {A.}~\bibnamefont
  {Togo}},\ }\href {http://phonondb.mtl.kyoto-u.ac.jp} {\bibinfo {title}
  {{Phonon Database at Kyoto University}}},\ \bibinfo {howpublished}
  {\url{http://phonondb.mtl.kyoto-u.ac.jp}}\BibitemShut {NoStop}%
\bibitem [{\citenamefont {Togo}\ and\ \citenamefont {Tanaka}(2015)}]{phonopy}%
  \BibitemOpen
  \bibfield  {author} {\bibinfo {author} {\bibfnamefont {A.}~\bibnamefont
  {Togo}}\ and\ \bibinfo {author} {\bibfnamefont {I.}~\bibnamefont {Tanaka}},\
  }\href {https://doi.org/10.1016/j.scriptamat.2015.07.021} {\bibfield
  {journal} {\bibinfo  {journal} {Scripta Materialia}\ }\textbf {\bibinfo
  {volume} {108}},\ \bibinfo {pages} {1} (\bibinfo {year} {2015})}\BibitemShut
  {NoStop}%
\bibitem [{\citenamefont {Helm}(1956)}]{Helm:1956zz}%
  \BibitemOpen
  \bibfield  {author} {\bibinfo {author} {\bibfnamefont {R.~H.}\ \bibnamefont
  {Helm}},\ }\href {https://doi.org/10.1103/PhysRev.104.1466} {\bibfield
  {journal} {\bibinfo  {journal} {Phys. Rev.}\ }\textbf {\bibinfo {volume}
  {104}},\ \bibinfo {pages} {1466} (\bibinfo {year} {1956})}\BibitemShut
  {NoStop}%
\end{thebibliography}%
\bibliographystyle{apsrev4-2}

\end{document}